\documentclass[aps,prd,twocolumn,amsmath,10pt,superscriptaddress,floatfix,nofootinbib]{revtex4-1}

\usepackage{epsfig,amssymb,amsfonts,amsmath,mathtools,bm,color,xcolor,graphicx}
\usepackage{comment}
\usepackage{hyperref}
\usepackage{multirow}
\hypersetup{pdftex,colorlinks=true,linkcolor=blue,citecolor=blue,menucolor=black,urlcolor=blue,filecolor=blue}

\usepackage{CJKutf8}

\begin{document}

\title{Faddeev equations for the $J/\psi\,NN$ and $\phi\,NN$ three-body systems in momentum space}

\author{Xu Zhang}
\affiliation{Institute of Theoretical Physics,
Chinese Academy of Sciences, Beijing 100190, China}
\email{zhangxu@itp.ac.cn}
\begin{abstract}
We construct Faddeev equations for the $J/\psi\,NN$ and $\phi\,NN$ three-body systems in momentum space. The two-body subsystem $t$-matrix is obtained by solving the Lippmann-Schwinger equations. The $NN$ interactions are constructed using the Malfliet-Tjon potential. The $J/\psi\,N$ interaction potentials in the ${}^4S_{3/2}$ 
and ${}^2S_{1/2}$ channels are taken from HAL QCD. The $\phi\,N$ interaction potential in the ${}^4S_{3/2}$ channel is also taken from HAL QCD. The ${}^2S_{1/2}$ $\phi N$ interaction is obtained from the combination of the $\phi p$ correlation function analysis and the HAL QCD results, as well as from the assumption that the spin-spin parts of the $\phi N$ and $J/\psi N$ interactions are inversely proportional to the respective hadron masses. The Faddeev equations are solved to find bound states in the $J/\psi\,NN$ and $\phi\,NN$ three-body systems. The numerical results suggest that there exist bound states in the $\phi\,NN$ three-body system, while there is no bound state in the $J/\psi\,NN$ system.

\end{abstract}


\maketitle

\section{Introduction}
\label{sec:intro}
Quantum chromodynamics (QCD) is the fundamental theory of the strong interaction, with quarks and gluons as degrees of freedom. However, first-principles calculations of hadron properties and interactions are challenging due to the nonperturbative nature of QCD at low energy scales~\cite{Gross:1973id,Politzer:1973fx,Gross:2022hyw}. Understanding the dynamics of the strong interaction at the hadronic scale has been important over the past several decades. The nucleon, which is a bound state of the strong interaction, is fundamentally described in terms of quarks and gluons in QCD. The $J/\psi$ has no valence components in common with the nucleon. Consequently, the dominant mechanism of the interaction between a nucleon and a $J/\psi$ at low energies may involve manifestations of the gluonic structure in the nucleon~\cite{Kharzeev:1995ij,Kharzeev:1998bz}.

The $c\bar c$ system is expected to have a small size $\propto 1/m_c$ compared to the nucleon. This suggests that the $J/\psi N\to J/\psi N$ reaction, driven by a gluonic van der Waals force, can be expressed in terms of gluonic matrix elements using the operator product expansion~\cite{Peskin:1979va,Bhanot:1979vb,Voloshin:1978hc,Gottfried:1977gp,Appelquist:1978rt,Luke:1992tm}. These processes provide crucial information about the trace anomaly contribution to the proton mass~\cite{Kharzeev:1995ij,Kharzeev:1998bz,Ji:1994av,Hatta:2018ina,Kou:2023zko}. Near the threshold, the relative momentum between the nucleon and the $J/\psi$ is small. The $J/\psi N$ dynamics mediated by multi-gluon exchange is expected to be attractive at short distances, and may generate a bound state between the $J/\psi$ and nucleons~\cite{Brodsky:1989jd,TarrusCastella:2018php}. In addition, the impact of the open-charm intermediate states on the $J/\psi N$ dynamics is investigated in Ref.~\cite{Wu:2024xwy}.
The hidden-charm pentaquark observed in the $J/\psi\, p$ spectrum at LHCb~\cite{LHCb:2015yax,LHCb:2016ztz,LHCb:2019kea} has a strong coupling to the open-charm states~\cite{Wu:2010jy,Liu:2019tjn,Fernandez-Ramirez:2019koa,Du:2019pij,Chen:2019bip,Chen:2019asm,He:2019ify,Guo:2019kdc,Xiao:2019mvs,Wang:2019ato,Xiao:2019aya,Meng:2019ilv,Voloshin:2019aut,Wang:2019hyc,Yamaguchi:2019seo,Lin:2019qiv,Gutsche:2019mkg,Burns:2019iih,Zhu:2019iwm,Wang:2019spc,Du:2021fmf,Shen:2024nck}. The contributions of the hidden-charm pentaquark to the $J/\psi N$ interaction also need to be understood properly.

Under the assumption of the vector meson dominance (VMD) model, the elastic $J/\psi\, p\to J/\psi\, p$ reaction is related to the $\gamma \,p\to J/\psi\, p$ reaction~\cite{Gell-Mann:1961jim,Kroll:1967it,Kuraev:1977fs,Balitsky:1978ic,Bauer:1977iq,Pumplin:1975fd,Barger:1975ng}. Near-threshold $J/\psi$ photoproduction provides an important process for analyzing the QCD dynamics of the charmonium--nucleon interaction. Early measurements were performed at Cornell~\cite{Gittelman:1975ix} and SLAC~\cite{Camerini:1975cy}. Recently, near-threshold $J/\psi$ photoproduction was measured by the GlueX Collaboration in Hall D at Jefferson Laboratory~\cite{GlueX:2019mkq,GlueX:2023pev}, the $J/\psi$-007 experiment in Hall C~\cite{Duran:2022xag,007:2026dow}, and the CLAS12 Collaboration in Hall B~\cite{CLAS:2026lls}. In Refs.~\cite{Strakovsky:2019bev,JointPhysicsAnalysisCenter:2023qgg,Strakovsky:2026qkv}, the $J/\psi\, p$ scattering length is extracted by analyzing the total or differential cross sections. Moreover, the effects of the hidden-charm pentaquark on the $J/\psi\, p$ spectrum in the photoproduction process were studied in Refs.~\cite{Wang:2015jsa,Kubarovsky:2015aaa,Karliner:2015voa,HillerBlin:2016odx,Winney:2019edt,Wang:2019krd,Wu:2019adv,Cao:2019kst,Paryev:2022wov,Strakovsky:2023kqu,Clymton:2024fbf,Duan:2024hby,Zhang:2024dkm,Clymton:2026kka}.

Lattice QCD has made progress in understanding the dynamics of the charmonium--nucleon interaction. Early lattice QCD calculations relied on the quenched approximation and the use of heavy pion masses~\cite{Yokokawa:2006td}. In Ref.~\cite{Skerbis:2018lew}, the energy levels of the $J/\psi N$ system were calculated in a finite box, and the results suggest that the scattering length tends to vanish. Recently, the interactions extracted using the HAL QCD method were found to be attractive at all distances and possess a long-range tail consistent with the two-pion exchange potential~\cite{Lyu:2024ttm}. The resulting $S$-wave scattering lengths are $0.30(2)\left(^{+0}_{-2}\right)$~fm and $0.38(4)\left(^{+0}_{-3}\right)$~fm for spin $3/2$ $J/\psi N$ and spin $1/2$ $J/\psi N$, respectively.

In addition to the two-body $J/\psi N$ system, the three-body $J/\psi NN$ system provides a promising window to understand the strong interaction. In the present work, the Faddeev equations in momentum space~\cite{Faddeev:1960su,faddeev:1965ml,Gloeckle1983} are constructed to investigate the three-body $J/\psi NN$ system. The two-body subsystem $t$-matrix is constructed using the Lippmann-Schwinger equations. The $NN$ interaction is built from the Malfliet-Tjon potential. The $J/\psi\,N$ interaction potentials in the ${}^4S_{3/2}$ 
and ${}^2S_{1/2}$ channels are taken from HAL QCD~\cite{Lyu:2024ttm}. 

The QCD van der Waals force may dominate the interaction between two hadrons when they have no common quarks, as suggested in Refs.~\cite{Peskin:1979va,Bhanot:1979vb,Voloshin:1978hc,Gottfried:1977gp,Appelquist:1978rt,Luke:1992tm,Brodsky:1989jd}. Analogously, assuming the $\phi$ meson is almost a pure $s\bar s$ state, the $\phi N$ interaction was expected to be dominated by the attractive QCD van der Waals force and to be strong enough to form a bound state~\cite{Gao:2000az,Gao:2017hya,Huang:2005gw,He:2018plt}. The $\phi N$ interaction has been studied in photoproduction at Spring8-LEPS~\cite{Ishikawa:2004id} and CLAS~\cite{CLAS:2010pxs,Dey:2014tfa}, and in proton-induced reactions at KEK-PS-E325~\cite{KEK-PS-E325:2005wbm} and COSY-ANKE~\cite{Polyanskiy:2010tj}. In Ref.~\cite{Strakovsky:2020uqs}, using the VMD model, the $\phi N$ scattering length was determined by analyzing the $\gamma\,p \to \phi\,p$ production cross section from the CLAS experiment~\cite{Dey:2014tfa}. Moreover, a comparison of the $\phi N$ and $J/\psi N$ scattering lengths will unveil the flavor-dependent effect in hadron physics and will improve our understanding of the strong interaction.

Recently, the $\phi N$ interaction was studied using the HAL QCD method at nearly physical quark masses~\cite{Lyu:2022imf}. In the $^4S_{3/2}$ channel, the open channels $\Lambda K(^2D_{3/2})$ and $\Sigma K(^2D_{3/2})$ are kinematically suppressed at low energies since those channels are $D$-wave. The $\phi N$ interaction in this channel is found to be a combination of a short-range attractive core and a long-range attractive tail. The resultant scattering length in the spin $3/2$ channel for $m_\pi=146.4$ MeV is $a_0^{(3/2)} = -1.43(23)_{\rm stat}\left(^{+36}_{-06}\right)_{\rm syst}$ fm. In the ${}^2S_{1/2}$ channel, the open channels $\Lambda K(^2S_{1/2})$ and $\Sigma K(^2S_{1/2})$ are $S$-wave. The $\phi N$ potential was found to have a strong coupling to the open channels~\cite{Lyu:2022imf}. The HAL QCD method does not provide reliable information on the potential in this channel.

Analogously, we will also study the three-body $\phi NN$ system using Faddeev equations in momentum space. Early studies of the three-body $\phi NN$ and $J/\psi NN$ systems using the Gaussian expansion method or Faddeev equations in configuration space can be found in Refs.~\cite{Belyaev:2007yc,Belyaev:2009ag,Sofianos_2010,Etminan:2024vkv,Filikhin:2024avj,Wen:2025wit,Lazauskas:2026nlg,Lazauskas:2026ysy,Yokota:2013sfa}. In this work, the inputs for the $\phi N$ interaction in the spin $3/2$ channel are taken from the HAL QCD results~\cite{Lyu:2022imf}. For the ${}^2S_{1/2}$ $\phi N$ interaction, we consider two schemes since there were no reliable lattice QCD results in this channel. 
In scheme $\rm I$, the ${}^2S_{1/2}$ $\phi N$ potential is a combination of the $\phi p$ correlation function analysis~\cite{Chizzali:2022pjd} and the HAL QCD results~\cite{Lyu:2022imf}, which was used in early studies of the $\phi NN$ three-body system~\cite{Wen:2025wit,Lazauskas:2026nlg,Lazauskas:2026ysy}.
In scheme $\rm II$, the ${}^2S_{1/2}$ $\phi N$ potential is connected to the  $\phi N(^4S_{3/2})$, $J/\psi N(^4S_{3/2})$, and $J/\psi N(^2S_{1/2})$ interactions, with the assumption that the spin-spin part of the $\phi N$ and $J/\psi N$ interactions is inversely proportional to the respective hadron masses~\cite{Wen:2025wit}.

The paper is organized as follows. In Section~\ref{sec:faddeev}, we construct the Faddeev equations for the $J/\psi\,NN$ and $\phi\,NN$ three-body systems in momentum space. In Section~\ref{sec:twobody}, the two-body interactions used in this work are described. In Section~\ref{sec:results}, the numerical results obtained by solving the Lippmann-Schwinger and the Faddeev equations are given. The summary is given in Section~\ref{sec:conclusions}.
\section{Faddeev equations in momentum space}
\label{sec:faddeev}   
\subsection{Faddeev decomposition}
\label{sec:decomp}
In this work, we will illustrate the momentum space treatment of the Faddeev equations~\cite{Faddeev:1960su,faddeev:1965ml,Gloeckle1983}. The Faddeev equations in configuration space can be found in Refs.~\cite{Laverne:1973kgl,Payne:1980ap,Payne:1980aq}. We consider the three-body $J/\psi NN$ and $\phi NN$ systems. The Hamiltonian of the three-body system can be written as
\begin{equation}
  H = H_0 + \sum_{i=1}^{3} V_i\,.
 \end{equation}
$H_0$ is the free kinetic energy operator of the three-body system. We denote the meson $J/\psi\,(\phi)$ as particle 1, and the two nucleons as particles 2 and 3. A potential $V_i$ is the two-body potential of the pair $(jk)$, with particle $i$ as spectator.

The Schr\"{o}dinger equation for the bound state at energy $E$
can be rewritten as
\begin{equation}
  |\Psi\rangle = G_0(E)\sum_{i=1}^{3}V_i|\Psi\rangle,
  \label{eq:schrodinger}
\end{equation}
where $G_0(E)=(E-H_0)^{-1}$ is the free three-body Green's function; $E$ is the total energy of the bound three-body system.

Following Refs.~\cite{Faddeev:1960su,faddeev:1965ml,Gloeckle1983}, the wave function can be decomposed into Faddeev amplitudes
\begin{equation}
  |\Psi\rangle = \sum_{i=1}^{3}|\psi_i\rangle,
  \label{eq:faddeevsum}
\end{equation}
where
\begin{equation}
  \label{eq:faddeevdef}
  |\psi_i\rangle = G_0(E)\,V_i|\Psi\rangle.
\end{equation}
Each amplitude $|\psi_i\rangle$ satisfies
\begin{equation}
  \label{eq:faddeevi}
  |\psi_i\rangle
  = G_0(E)\,V_i\sum_{j=1}^3|\psi_j\rangle.
\end{equation}
Since two nucleons are identical, quantities that carry different particle indices, such as operators and states, can be related to each other simply by means of permutations, e.g.,
\begin{equation}
V_3 = P_{23}\,V_2\,P_{23}^{-1}, 
\end{equation}
where $P_{23}$ is the transposition of particles 2 and 3. Imposing antisymmetry, the Faddeev amplitude transforms as
\begin{equation}
  -|\psi_3\rangle =P_{23} |\psi_2\rangle.
  \label{eq:psi3}
\end{equation}
Thus, the wave function takes the form
\begin{equation}
  \label{eq:wavefn}
  |\Psi\rangle = |\psi_1\rangle + (1-P_{23})|\psi_2\rangle.
\end{equation}
Inserting Eq.~\eqref{eq:wavefn} into Eq.~\eqref{eq:faddeevi}, one can get
\begin{align}
   |\psi_1\rangle &= G_0(E)\,V_1\,\Big(|\psi_1\rangle+(1-P_{23})\,|\psi_2\rangle \Big),\nonumber \\
  |\psi_2\rangle  &= G_0(E)\,V_2\,\Big(|\psi_1\rangle+(1-P_{23})\,|\psi_2\rangle \Big).
\end{align}
Moving $G_0(E)\,V_i\,|\psi_i\rangle$ to the left side and multiplying the resulting equations with $[1-G_0(E)\,V_i\,]^{-1}$ from the left yields
\begin{align}
   |\psi_1\rangle &=[1-G_0(E)\,V_1\,]^{-1} G_0(E)\,V_1\,\Big((1-P_{23})\,|\psi_2\rangle \Big),\nonumber \\
  |\psi_2\rangle  &=[1-G_0(E)\,V_2\,]^{-1} G_0(E)\,V_2\,\Big(|\psi_1\rangle-P_{23}\,|\psi_2\rangle \Big).
\end{align}
The factor $[1-G_0(E)\,V_i\,]^{-1} G_0(E)$ is the three-body Green's function $G_i(E)$ of channel $i$ with the two-body interaction $V_i$,
\begin{align}
  G_i(E)& = (E - H_0 - V_i)^{-1} \nonumber \\
 & = G_0(E) + G_0(E)\,T_i(E)\,G_0(E).
\end{align}
$T_i(E)$ is the two-body transition matrix in the three-body space, given by the Lippmann-Schwinger equation
\begin{equation}
  T_i(E) = V_i + V_i\,G_0(E)\,T_i(E).
\end{equation}
Using the relation $G_i(E)\,V_i = G_0(E)\,T_i(E)$, the Faddeev amplitudes can be written as
\begin{align}
\label{eq:faddeevamp}
   |\psi_1\rangle &=G_0(E)\,T_1(E)\,\Big((1-P_{23})\,|\psi_2\rangle \Big),\nonumber \\
  |\psi_2\rangle  &=G_0(E)\,T_2(E)\,\Big(|\psi_1\rangle-P_{23}\,|\psi_2\rangle \Big).
\end{align}

\subsection{Partial-wave expansion}
\label{sec:pwbasis}
We have three particles with masses $m_1$, $m_2$, $m_3$ and momenta $\bm k_1$, $\bm k_2$, $\bm k_3$. The relative momenta $\bm{p}_{i}$ and $\bm{q}_{i}$ are defined as
\begin{align}
\bm{p}_{i} &= \frac{m_{k}\bm{k}_{j} - m_{j}\bm{k}_{k}}{m_{jk}}, \nonumber \\
\bm{q}_{i} &= \frac{m_{jk}\bm{k}_{i}-m_{i}(\bm{k}_{j} + \bm{k}_{k}) }{m_{ijk}},
\end{align}
where $m_{ij}=m_i+m_j$, $m_{ijk}=m_i+m_j+m_k$, and $(i,j,k)$ takes the values $(1,2,3)$, $(2,3,1)$, $(3,1,2)$. Related to them are the momentum states $|\bm{p}_i\bm q_i\rangle$, which are normalized as
\begin{equation}
 \langle\bm{p}_i\bm q_i|\bm{p}_i'\bm q_i'\rangle =  \delta(\bm{p}_i-\bm{p}_i')\delta(\bm{q}_i-\bm{q}_i').
\end{equation}
We introduce the partial-wave-projected states $|p\,lm_l\rangle$ and $|q\,\lambda m_{\lambda}\rangle$, and both are defined as
\begin{equation}
\langle \bm k'|k\,lm\rangle=\frac{\delta(k-k')}{kk'}\mathcal{Y}_{lm}(\hat{k}').
\end{equation}
Then we can form states of total orbital angular momentum $L$,
\begin{equation}
| pq(l\lambda)LM\rangle=\sum_{m_l m_{\lambda}} (l\lambda L,m_l m_{\lambda} M) |p\,lm_l\rangle|q\,\lambda m_{\lambda}\rangle,
\end{equation}
where $l$ denotes the orbital angular momentum, and $\lambda$ is the orbital angular momentum of the spectator relative to the c.m. of the pair. The relation between the partial-wave state and the momentum state is
\begin{equation}
\langle \bm p'\bm q'\mid pq(l\lambda)LM\rangle
\equiv \frac{\delta(p-p')}{pp'}\,\frac{\delta(q-q')}{qq'}\,
\mathcal{Y}_{l\lambda}^{LM}(\hat{\bm{p}}'\hat{\bm{q}}'),
\end{equation}
where $\mathcal{Y}_{l\lambda}^{LM}(\hat{\bm{p}}'\hat{\bm{q}}')$ is the simultaneous eigenfunction of $l^2$, $\lambda^2$, $L^2$, and $L_z$~\cite{BlattWeisskopf1952}.

In the $LS$-coupling basis, taking into account the spin and isospin, the partial-wave state is defined as
\begin{equation}
|pq\alpha \rangle_{i} = |p_i q_i; (l\lambda)L (s s_i) S (LS)J (tt_i) {T} \rangle.
\end{equation}
The subscript $i$ indicates that particle $i$ is a spectator and the particles $j$ and $k$ form a pair. The quantum numbers $s$ and $t$ denote the total spin and isospin of the pair. The quantum numbers $s_i$ and $t_i$ are the spin and isospin of the spectator. The total three-particle angular momentum and total isospin are denoted by $J$ and $T$, respectively.

Moreover, we can combine $l$ and $s$ into the total subsystem angular momentum $j$, and $\lambda$ and $s_i$ into the total spectator angular momentum $I$, which finally combine to form $J$. The basis of the $jI$-coupling is related to the basis of the $LS$-coupling as
\begin{align}
&| p_i q_i; (ls)j (\lambda s_i) I (jI)J (tt_i) {T}\rangle = \sum_{LS}
  \sqrt{\hat{j}\,\hat{I}\,\hat{L}\,\hat{S}}\;
  \nonumber \\
 &\times \begin{Bmatrix}l & s & j \\ \lambda & s_i & I \\ L & S & J\end{Bmatrix} | p_i q_i; (l\lambda)L (s s_i) S (LS)J (tt_i) {T} \rangle.
\end{align}
We have introduced the convenient abbreviation $\hat{l}=2l+1$. We denote the quantum numbers in the basis of $jI$-coupling by $\alpha$, and the corresponding basis state by $|pq\alpha\rangle$.

The inputs of the Faddeev amplitudes need the off-shell two-body transition amplitude ${}_{i}\!\langle pq\alpha|T_i(E)|p'q'\alpha'\rangle_{i}$ in three-body space. The transition amplitude in three-body space is related to the corresponding amplitude in two-body space as
\begin{align}
 \label{eq:tmatrix}
 & {}_{i}\!\langle pq\alpha|T_i(E)|p'q'\alpha'\rangle_{i} =\,\delta_{JJ'}\,\delta_{TT'}\,\delta_{\lambda\lambda'}\,\delta_{II'}\,\delta_{tt'}\,\delta_{jj'}\nonumber \\
 & \times \frac{\delta(q-q')}{q^2}\,
    t_{i,nn'}(p,p',E - q^2/(2\mu_i)),
 \end{align}
where $n$ denotes the subset of pair quantum numbers $l,s,j$ and $t$.
The reduced masses $\mu_i$ of the spectator $i$ and
the pair $(jk)$ are defined as
\begin{equation}
\mu_{i} = \frac{m_i(m_j+m_k)}{m_{ijk}}.
\end{equation}
The two-body partial-wave amplitude $t_{i,nn'}$ satisfies the 
Lippmann-Schwinger equation,
\begin{align}
 \label{eq:ls}
  t_{i,nn'}(p,&\,p',E) = V_{i,nn'}(p,p',E) \nonumber \\
 &  + \sum_{n''}\int_0^\infty\!dk\,k^2\,
    \frac{V_{i,nn''}(p,k,E)\,t_{i,n''n'}(k,p',E)}
         {E - k^2/(2\nu_i)},
\end{align}
where $\nu_i$ is the reduced mass of particles $j$ and $k$,
\begin{align}
  \nu_i = \frac{m_j\,m_k}{m_{jk}}.   
\end{align}
In this work, the two-body Lippmann-Schwinger equation is solved using the matrix inversion method~\cite{Haftel:1970zz}.
The free three-body propagator is
\begin{align}
  \label{eq:G0}
  {}_{i}\!\langle pq\alpha|G_0(E)|p'q'\alpha'&\rangle_{i}
  = \delta_{\alpha\alpha'}\,
    \frac{\delta(q-q')}{q^2}\,
    \frac{\delta(p-p')}{p^2}\notag\\
  &\quad\times
    \frac{1}{E - p^2/(2\nu_i) - q^2/(2\mu_i)}.
\end{align}
We are left with the geometrical quantities, which link different arrangements of three
particles. The matrix element can be written as
\begin{widetext}
\begin{align}
\label{eq:geome}
 {}_{i}\langle pq\alpha\mid p'q'\alpha'\rangle_{j}\, 
 =\int_{-1}^1 dx  
       \frac{\delta(p-\pi_{ij}(q,q',x))}{p^{l+2}}\;
  \frac{\delta(p'-\pi_{ij}'(q,q',x))}{p'^{l'+2}} G_{\alpha\alpha'}^{ij}(q,q',x), 
\end{align}
where
\begin{align}
\label{eq:pipi}
 \pi_{ij}(q,q',x)=\sqrt{\tfrac{m_j^2}{m_{jk}^2}q^2+q'^2+\tfrac{2m_j}{m_{jk}}qq'x}\,,\quad
  \pi_{ij}'(q,q',x)=\sqrt{q^2+\tfrac{m_i^2}{m_{ik}^2}q'^2+\tfrac{2m_i}{m_{ik}}qq'x}\,.
\end{align}
The calculation of the geometrical quantities $G_{\alpha\alpha'}^{ij}(q,q',x)$ in Eq.~\eqref{eq:geome} is tedious. As an example, we show the matrix element ${}_{1}\langle pq\alpha\mid p'q'\alpha'\rangle_{2}$ in the Appendix~\ref{app:permuope}, and other matrix elements are analogous.
  
Inserting the operator representations in Eqs.~\eqref{eq:tmatrix}, \eqref{eq:G0} and \eqref{eq:geome} into Eq.~\eqref{eq:faddeevamp}, and projecting into partial-wave states, the Faddeev equations take the integral form~\cite{Gloeckle1983},
\begin{align}
  \label{eq:faddeevmomen1}
  \psi_1(p,q,\alpha)
  &= \frac{1}{E-p^2/(2\nu_1)-q^2/(2\mu_1)} \sum_{\alpha'\alpha''}
  \int_0^\infty\!q'^2\,dq'\,
  \int_0^\infty\!q''^2\,dq''\int_{-1}^{1}\!dx\; {}_{1}\!\langle pq\alpha|T_1(E)|\pi_{12}q'\alpha'\rangle_{1}\nonumber \\
 &\times 2 \frac{G_{\alpha'\alpha''}^{12}(q',q'',x)}{\pi_{12}^{l'}(q',q'',x)\pi_{12}'^{l''}(q',q'',x)} \psi_2(\pi_{12}',q'',\alpha'')\,,\\
 \label{eq:faddeevmomen2}
 \psi_2(p,q,\alpha)
  &= \frac{1}{E-p^2/(2\nu_2)-q^2/(2\mu_2)} \sum_{\alpha'\alpha''}
  \int_0^\infty\!q'^2\,dq'\,
  \int_0^\infty\!q''^2\,dq''\int_{-1}^{1}\!dx\; {}_{2}\!\langle pq\alpha|T_2(E)|\pi_{12}q'\alpha'\rangle_{2}\nonumber \\
 &\times\Big( \frac{G_{\alpha'\alpha''}^{21}(q',q'',x)}{\pi_{21}^{l'}(q',q'',x)\pi_{21}'^{l''}(q',q'',x)} \psi_1(\pi_{21}',q'',\alpha'') +\frac{G_{\alpha'\alpha''}^{23}(q',q'',x)}{\pi_{23}^{l'}(q',q'',x)\pi_{23}'^{l''}(q',q'',x)} \psi_2(\pi_{23}',q'',\alpha'')  \Big)\,.
\end{align}
\end{widetext}
This is a homogeneous algebraic set of equations for the unknown Faddeev components $\psi_i(p,q,\alpha)$.
The integral equations couple partial waves. 
The summations $\alpha'$ and $\alpha''$ are over the three-body partial-wave states. In this work, only $S$-wave interactions are considered.

The Faddeev equations in Eqs.~\eqref{eq:faddeevmomen1} and \eqref{eq:faddeevmomen2} have 
the general structure
\begin{equation}
 \label{eq:eigenvalue}
 \bm K(E)\,|\psi\rangle = \,|\psi\rangle,
\end{equation}
where $\bm K(E)$ is the kernel of the Faddeev equations. It is useful to consider the auxiliary  eigenvalue problem
\begin{equation}
 \label{eq:eigenlua}
 \bm K(z)\,|\varphi(z)\rangle = \lambda(z)\,|\varphi(z)\rangle,
\end{equation}
where $\lambda(z)$ denotes the eigenvalues of the operator $\bm K(z)$ and  $|\varphi(z)\rangle$ the corresponding eigenvectors. The strategy for solving Eq.~\eqref{eq:eigenlua} is to find the value $z=E$ such that the eigenvalue $\lambda(E)=1$. Then $E$ is the physical binding energy and the corresponding eigenvector $|\varphi(E)\rangle$ is the Faddeev amplitude.

In this work, the Faddeev equations are solved by discretizing the equations into a matrix form and then diagonalizing the matrix. The Faddeev components under the integral in Eq.~\eqref{eq:faddeevmomen1} can be written in the form
\begin{equation}
 \psi_2(\pi_{12}',q'',\alpha'') =\sum_k S_k(\pi_{12}')  \psi_2(p_k,q'',\alpha''), 
\end{equation}
which is based on cubic splines~\cite{BurdenFaires2016,
PressTeukolsky2007}. The same treatment applies to the Faddeev components under the integral in Eq.~\eqref{eq:faddeevmomen2}. The set of $p_k$ is chosen sufficiently dense to guarantee the desired quality of interpolation. Then using Gauss-Legendre integration, the Faddeev equations in Eqs.~\eqref{eq:faddeevmomen1} and \eqref{eq:faddeevmomen2} are transformed into a system of linear homogeneous equations in a discretised form. Other numerical methods based on iteration can be found in Refs.~\cite{Stadler:1991zz,Gloeckle:1982ze,Chen:1986pi,Chen:1991zza}.

\section{Two-Body Interactions}
\label{sec:twobody}

\subsection{$NN$ interaction}
\label{sec:nnpote}

Over several decades, considerable effort has been made to develop the realistic 
$NN$ interaction. Phenomenological interactions such as the Argonne V18~\cite{Wiringa:1994wb}, the charge-dependent Bonn potential~\cite{Machleidt:2000ge}, the Nijmegen potential~\cite{Stoks:1994wp}, along with the chiral effective field theory~\cite{Epelbaum:2014efa} emerging from low-energy QCD have achieved excellent progress in describing $NN$ scattering data.  In this work, we adopt the Malfliet-Tjon (MT) potential~\cite{Malfliet:1968tj,Friar:1990zza}, which allows a direct comparison with earlier investigations of the $J/\psi NN$ and $\phi NN$ systems~\cite{Wen:2025wit,Lazauskas:2026nlg,Lazauskas:2026ysy}.

The interaction potential is constructed as a sum of a repulsive and an attractive Yukawa term,
\begin{align}
  \label{eq:vmt}
 & V_{NN}^{{}^1S_0}(r) = C_1^{{}^1S_0}\,\frac{e^{-\mu_1^{MT} r}}{r}
                + C_2^{{}^1S_0}\,\frac{e^{-\mu_2^{MT} r}}{r}, \nonumber \\
 & V_{NN}^{{}^3S_1}(r) =C_1^{{}^3S_1}\, \frac{e^{-\mu_1^{MT} r}}{r}
                + C_2^{{}^3S_1}\,\frac{e^{-\mu_2^{MT} r}}{r}.              
\end{align}
The parameters are given in Table~\ref{tab:NNpar}. Despite its simplicity, this
model reproduces reasonably well the binding energies of
two-, three-, and four-nucleon systems~\cite{Friar:1990zza,Gloeckle:1995jg,Navratil:1999pw,Kievsky:2008es,Leidemann:2012hr}.

\begin{table}[htb]
\caption{The parameters of the MT $NN$ potential given in Eq.~\eqref{eq:vmt} for the singlet ${}^1S_0$ and triplet ${}^3S_1$ channels.}
\label{tab:NNpar}
\begin{ruledtabular}
\begin{tabular}{ccccc}
Channels & $C_1$[MeV $\cdot$ fm] & $\mu_1^{MT}$[fm$^{-1}$] & $C_2$[MeV $\cdot$ fm] & $\mu_2^{MT}$[fm$^{-1}$] \\
\hline
${}^1S_0$ & $-514$  & $1.55$ & $1439$ & $3.11$ \\
${}^3S_1$ & $-627$  & $1.55$ & $1439$ & $3.11$ 
\end{tabular}
\end{ruledtabular}
\end{table}

The transition between configuration space and momentum space for a fixed $l$ is
\begin{align}
V_\ell (p, p') &= \int d\hat{p} \mathcal{Y}_{\ell m}^* (\hat{p}) \frac{1}{(2\pi)^{3/2}} \int d\bm{r} e^{i\bm{p} \cdot \bm{r}} V(r) \nonumber \\
&\quad \times \int d\hat{p}' \frac{1}{(2\pi)^{3/2}} e^{-i\bm{p}' \cdot \bm{r}} \mathcal{Y}_{\ell m} (\hat{p}') \nonumber \\
&= \frac{2}{\pi} \int_0^\infty dr \, r^2 j_\ell (pr) V(r) j_\ell (p'r),
\end{align}
where $j_\ell (pr)$ is the spherical Bessel function~\cite{abramowitz+stegun}.

\subsection{$J/\psi N$ interaction}
\label{sec:jpsinpote}

The $J/\psi N$ interaction potential in the ${}^4S_{3/2}$ and ${}^2S_{1/2}$ channels is taken from the HAL QCD analysis in Ref.~\cite{Lyu:2024ttm}. They performed an uncorrelated fit with a phenomenological three-range Gaussian function,
\begin{align}
\label{eq:psinqt}
&V_{J/\psi N}^{{}^4S_{3/2}}(r)=-\sum_{i=1}^{3}a_i^{{}^4S_{3/2}}
    e^{-\,({r}/{b_i^{{}^4S_{3/2}}})^2},\nonumber \\
&V_{J/\psi N}^{{}^2S_{1/2}}(r)=-\sum_{i=1}^{3}a_i^{{}^2S_{1/2}}
    e^{-\,({r}/{b_i^{{}^2S_{1/2}}})^2}.
\end{align}
The parameters are given in Table~\ref{tab:jpsin}.
\begin{table}[htb]
\caption{The parameters of the $J/\psi N$ potential given in Eq.~\eqref{eq:psinqt}. The values are taken from the HAL QCD analysis in Ref.~\cite{Lyu:2024ttm}. $a_1$, $a_2$ and $a_3$ are in units of MeV. $b_1$, $b_2$ and $b_3$ are in units of fm.}
\label{tab:jpsin}
\begin{ruledtabular}
\begin{tabular}{ccccccc}
Channels & $a_1$  & $b_1$ & $a_2$& $b_2$ & $a_3$ & $b_3$  \\
\hline
${}^4S_{3/2}$ & $51(1)$  & $0.09(1)$ & $13(6)$ & $0.49(7)$& $22(5)$ & $0.82(6)$ \\
${}^2S_{1/2}$ & $101(1)$  & $0.13(1)$ & $33(6)$ & $0.44(5)$& $23(8)$ & $0.83(9)$
\end{tabular}
\end{ruledtabular}
\end{table}

\subsection{$\phi N$ interaction}
The $\phi N$ interaction has been studied using the HAL QCD method near physical quark masses~\cite{Lyu:2022imf}. They suggested that the $\phi N$ correlation function in the ${}^4S_{3/2}$ channel is dominated by the elastic scattering states with no significant effects from the two-body open channels $\Lambda K(^2D_{3/2})$, $\Sigma K(^2D_{3/2})$, and the three-body open channels including $N\phi \to \{\Sigma^* K, \Lambda(1405) K\} \to \{\Lambda \pi K, \Sigma \pi K\}$. In contrast, the $\phi N$ potential in the ${}^2S_{1/2}$ channel has a strong coupling to the open channels $\Lambda K(^2S_{1/2})$ and $\Sigma K(^2S_{1/2})$, which makes the information on the potential in this channel from lattice QCD have large uncertainties, as discussed in Ref.~\cite{Lyu:2022imf}.

The $\phi N$ interaction potential in the ${}^4S_{3/2}$ channel from HAL QCD is a sum
of two Gaussian terms to describe the attractive short-range part, together with a two-pion exchange tail at large distances,
\begin{align}
 \label{eq:phiquartet}
  V_{\phi N}^{{}^4S_{3/2}}(r)
  &= \alpha_1 e^{-(r/{\beta}_1)^2}
   + \alpha_2 e^{-(r/\beta_2)^2}\notag\\
  &\quad +\alpha_3 m_\pi^4\,f(r,\beta_3)
    \frac{e^{-2m_\pi r}}{r^2},
\end{align}
with the Argonne-type form factor~\cite{Wiringa:1994wb},
\begin{align}
f(r,\beta_3)=(1-{e^{-(r/\beta_3)^2}})^2.
\end{align}
The parameters are given in Table~\ref{tab:phinqt}.
\begin{table}[htb]
\caption{Parameters of the $\phi N$ potential in the ${}^4S_{3/2}$ channel in Eq.~\eqref{eq:phiquartet}. The values are taken from the HAL QCD analysis in Ref.~\cite{Lyu:2022imf}.
$\alpha_1$, $\alpha_2$ are in units of MeV. $\beta_1$, $\beta_2$, $\beta_3$ and $\alpha_3m_{\pi}^{4}$ are in units of fm.}
\label{tab:phinqt}
\begin{ruledtabular}
\begin{tabular}{ccccccc}
 $\alpha_1$  & $\beta_1$ & $\alpha_2$& $\beta_2$ & $\alpha_3m_{\pi}^{4}$ & $\beta_3$  \\
\hline
$-371(27)$ & $0.13(1)$  & $-119(39)$ & $0.30(5)$ & $-97(14)$& $0.63(4)$ 
\end{tabular}
\end{ruledtabular}
\end{table}

In the ${}^2S_{1/2}$ channel, the HAL QCD results have large uncertainties. We will consider two schemes. In scheme $\rm I$, the ${}^2S_{1/2}$ $\phi N$ potential is the combination of the $\phi p$ correlation function analysis~\cite{Chizzali:2022pjd} and the HAL QCD results~\cite{Lyu:2022imf}. The long-range two-pion exchange potential should also characterize the ${}^2S_{1/2}$ channel as it does in the ${}^4S_{3/2}$ channel, since the exchange of two pions in a scalar-isoscalar state, such as $\sigma$, does not depend on the total spin of the $\phi N$ system. The $\phi N$ potential in the ${}^2S_{1/2}$ channel is assumed to be a sum of two Gaussian terms modified with an additional factor, together with a two-pion exchange tail. Then the potential is written as
\begin{align}
  V_{\phi N}^{{}^2S_{1/2}}(r)
  =& \beta\Big(\alpha_1 e^{-(r/\beta_1)^2}
   + \alpha_2 e^{-(r/\beta_2)^2}\Big)\nonumber\\
  &+\alpha_3 m_\pi^4\,f(r,\beta_3)\frac{e^{-2m_\pi r}}{r^2}.
\end{align}
The values of $\alpha_1$, $\alpha_2$, $\alpha_3$, $\beta_1$, $\beta_2$, and $\beta_3$ are the same as those in the ${}^4S_{3/2}$ channel from HAL QCD. The value $\beta=6.9$ is determined by fitting the $\phi p$ correlation function measured by ALICE, as discussed in Ref.~\cite{Chizzali:2022pjd}.

In scheme $\rm II$, the ${}^2S_{1/2}$ $\phi N$ potential is constructed with the assumption that the spin-spin parts of the $J/\psi N$ and $\phi N$ interactions are inversely proportional to the hadron masses. Then the ${}^2S_{1/2}$ $\phi N$ potential can be obtained as~\cite{Wen:2025wit}
\begin{align}
V^{Spin}_{J/\psi N}(r)=&\frac{V^{{}^4S_{3/2}}_{J/\psi N}-V^{{}^2S_{1/2}}_{J/\psi N}}{3}  , \label{eq:phiNBspin}\\
V^{Spin}_{\phi N}(r)=&\frac{m_{J/\psi}}{m_\phi}V^{Spin}_{J/\psi N}(r),\\
V^{{}^2S_{1/2}}_{\phi N}(r)=&V^{{}^4S_{3/2}}_{\phi N}(r)-3 V^{Spin}_{\phi N}(r) .\label{eq:phiN1B}
\end{align}
The spin-spin interaction is labeled by the superscript ``Spin''.
\section{Numerical Results}
\label{sec:results}

\subsection{Two-body system}
First, we consider the two-body system. The two-body results are 
obtained by solving the Lippmann-Schwinger equation in Eq.~\eqref{eq:ls}.
The calculated two-body binding energy $B_2$, scattering length $a$, and effective range $r_0$ for $NN$, $J/\psi N$, and $\phi N$ are shown in Table~\ref{tab:twobyres}.
We have considered the experimental masses $m_{\pi}=138$, $m_{\phi}=1019$, and $m_{N}=939$~MeV, and the masses used in HAL QCD, $m_{\pi}=146$, $m_{\phi}=1048$, and $m_{N}=954$~MeV. The mass of $J/\psi$ is taken to be $m_{J/\psi}=3096$~MeV.

\begin{table}[htb]
\caption{The two-body results obtained by solving the Lippmann-Schwinger equation in Eq.~\eqref{eq:ls}. Results correspond to experimental masses and the masses used in HAL QCD (in parentheses).}
\label{tab:twobyres}
\begin{ruledtabular}
\begin{tabular}{cccc}
Channels & $B_2$~[MeV] & $a_0$~[fm] & $r_0$~[fm]\\
\hline
$NN({}^1S_0)$& - & $-23.57(-29.94)$ & 2.90(2.86) \\
$NN({}^3S_1)$& $2.23(2.41)$ & $5.51(5.32)$& 1.88(1.86)  \\
$J/\psi N({}^4S_{3/2})$ & -& -0.29(-0.30) & 3.24(3.21) \\
$J/\psi N({}^2S_{1/2})$ & -& -0.37(-0.38) & 2.64(2.62)\\
$\phi N({}^4S_{3/2})$ & - & -1.66(-1.45) & 2.36(2.36) \\
$\phi N({}^2S_{1/2})$, $\rm I$ & 16.61(18.41) & 1.87(1.76) & 0.47(0.44) \\
$\phi N({}^2S_{1/2})$, $\rm II$ & -&-2.57(-2.22) & 1.90(1.89) \\
\end{tabular}
\end{ruledtabular}
\end{table}

When we use the physical nucleon masses, the MT potential reproduces the deuteron
with binding energy $2.23$~MeV. This parametrization yields the scattering length $a_0=5.51$ fm, 
effective range $r_0=1.88$ fm in the ${}^3S_1$ channel, and $a_0=-23.57$ fm, 
effective range $r_0=2.90$ fm in the ${}^1S_0$ channel. We can find that the physical masses reproduce fairly well the deuteron energy, the scattering length, and the effective range. When we use the masses employed in HAL QCD, the results are slightly modified.

Using the HAL QCD potentials, the $J/\psi N$ interaction in neither the ${}^4S_{3/2}$ nor the ${}^2S_{1/2}$ channel generates a bound state. The $J/\psi N$ scattering length is rather small. Using the masses employed in HAL QCD, the Lippmann-Schwinger equation reproduces the scattering length extracted by HAL QCD for the $J/\psi N$ interaction~\cite{Lyu:2024ttm}.

The $\phi N$ interaction in the ${}^4S_{3/2}$ channel does not form a bound state using the HAL QCD potential~\cite{Lyu:2022imf}. Using the masses employed in HAL QCD, the Lippmann-Schwinger equation reproduces the scattering length and effective range extracted from HAL QCD in this channel~\cite{Lyu:2022imf}. The $\phi N$ interaction in ${}^2S_{1/2}$ channel, in scheme $\rm I$, is strong enough to produce a bound state with binding energy around $20$~MeV.

\subsection{$J/\psi\,NN$ three-body system}

In this work, the Faddeev equations are solved by performing matrix diagonalization
to obtain the eigenvalues. We did not find a bound state in the $J/\psi\,NN$ three-body system. The $J/\psi N$ scattering lengths in the ${}^4S_{3/2}$ and ${}^2S_{1/2}$ channels are smaller than the corresponding $\phi N$ scattering lengths by one order of magnitude. The $J/\psi N$ interaction is very weak, and as a result, neither a two-body $J/\psi N$ bound state nor a three-body $J/\psi NN$ bound state is formed.
Our results are consistent with the conclusions obtained using the Gaussian expansion method in Refs.~\cite{Wen:2025wit,Yokota:2013sfa}.

\subsection{$\phi\,NN$ three-body system}

The calculated binding energies for the $\phi NN$ three-body system corresponding to schemes $\rm I$ and $\rm II$ are shown in Tables~\ref{tab:phiA} and \ref{tab:phiB}, respectively. The binding energies are relative to the three-body threshold.

\subsubsection{Scheme $\rm I$}

In scheme $\rm I$, there exists a deep bound state in the $(I)J^P=(0)0^-$ channel. In this channel, the two-body subsystem 
$NN$ is in the ${}^3S_1$ channel, and $\phi N$ is in the ${}^2S_{1/2}$ channel. 
The $NN$ interaction in the ${}^3S_1$ channel is strong and forms the deuteron. The $\phi N$ interaction in the ${}^2S_{1/2}$ channel is even stronger, and can form a bound state with binding energy around $20$~MeV. Both the strong $NN$ interaction in the ${}^3S_1$ channel and the $\phi N$ interaction in the ${}^2S_{1/2}$ channel lead to a deep three-body $\phi NN$ bound state in the $(I)J^P=(0)0^-$ channel. Moreover, we can find that using the masses employed in HAL QCD leads to a slight modification of $B_3$, compared to the results using the experimental masses. The numerical values obtained in this work agree with those from Refs.~\cite{Wen:2025wit,Lazauskas:2026nlg,Lazauskas:2026ysy}.

\begin{table}[!htb]
\caption{The $\phi\,NN$ three-body binding energy in scheme $\rm I$. Results correspond to experimental masses and the masses used in HAL QCD (in parentheses). The binding energies are relative to the three-body threshold.}
\label{tab:phiA}
\begin{ruledtabular}
\begin{tabular}{ccc}
Channel & $B_3$~[MeV] \\
\hline
$I=0,\,J^P=0^-$ & $50.25(52.45)$  \\
$I=0,\,J^P=1^-$ & $32.13(33.69)$  \\
$I=1,\,J^P=1^-$ & -(-)   \\
$I=0,\,J^P=2^-$ & $2.39(2.45)$  \\
\end{tabular}
\end{ruledtabular}
\end{table}
\begin{table}[!htb]
\caption{The $\phi\,NN$ three-body binding energy in scheme $\rm II$. Results correspond to experimental masses and the masses used in HAL QCD (in parentheses). The binding energies are relative to the three-body threshold.}
\label{tab:phiB}
\begin{ruledtabular}
\begin{tabular}{ccc}
Channel & $B_3$~[MeV] \\
\hline
$I=0,\,J^P=0^-$ & $3.57(3.38)$  \\
$I=0,\,J^P=1^-$ & $3.06(2.95)$  \\
$I=1,\,J^P=1^-$ & - (-)   \\
$I=0,\,J^P=2^-$ & $2.39(2.45)$  \\
\end{tabular}
\end{ruledtabular}
\end{table}

In the $(I)J^P=(0)2^-$ channel, there exists a shallow bound state in the three-body $\phi NN$ system. In this channel, the two-body subsystem $NN$ is in the ${}^3S_1$ channel, and $\phi N$ is in the ${}^4S_{3/2}$ channel. The strength of the subsystem $NN$ interaction in the $(I)J^P=(0)2^-$ channel is the same as that in the $(I)J^P=(0)0^-$ channel. The $\phi N$ interaction in the ${}^4S_{3/2}$ channel cannot form a bound state, and is weaker than that in the ${}^2S_{1/2}$ channel. Then the interactions of $NN$ in the ${}^3S_1$ channel and $\phi N$ in the ${}^4S_{3/2}$ channel lead to a three-body $\phi NN$ bound state in the $(I)J^P=(0)2^-$ channel, whose binding energy is much smaller than that in the $(I)J^P=(0)0^-$ channel.

In the $(I)J^P=(0)1^-$ channel, there exists a bound state in the three-body $\phi NN$ system. In this channel, the two-body subsystem $NN$ is in the ${}^3S_1$ channel, and the $\phi N$ interaction is a mixture of the ${}^2S_{1/2}$ and the ${}^4S_{3/2}$ channels. The strength of the subsystem $NN$ interaction in the $(I)J^P=(0)1^-$ channel is the same as that in the $(I)J^P=(0)0^-$ and $(0)2^-$ channels. Then the $NN$ interaction in the ${}^3S_1$ channel and a mixture of the ${}^2S_{1/2}$ and the ${}^4S_{3/2}$ $\phi N$ interactions lead to a three-body $\phi NN$ bound state in the $(I)J^P=(0)1^-$ channel. The binding energy of the bound state in this channel lies between those in the $(I)J^P=(0)0^-$ and the $(I)J^P=(0)2^-$ channels.

In the $(I)J^P=(1)1^-$ channel, we did not find a bound state in the three-body $\phi NN$ system. In this channel, the two-body subsystem $NN$ is in the ${}^1S_0$ channel. The $NN$ interaction in the ${}^1S_0$ channel forms a virtual state, and the strength of the interaction is weaker than that in the ${}^3S_1$ channel. In the $(I)J^P=(1)1^-$ channel, the $\phi N$ interaction is also a mixture of the ${}^2S_{1/2}$ and the ${}^4S_{3/2}$ channels. However, the mixture in this channel is different from that in the $(I)J^P=(0)1^-$ channel according to the geometrical quantities in Eq.~\eqref{eq:geome}.

\subsubsection{Scheme $\rm II$}

In scheme $\rm II$, the strength of the ${}^2S_{1/2}$ $\phi N$ interaction is weaker compared to scheme $\rm I$.
The ${}^2S_{1/2}$ $\phi N$ interaction cannot form a two-body bound state. We found bound states in the three-body $\phi NN$ system in the $(I)J^P=(0)0^-$, $(0)1^-$, and $(0)2^-$ channels. In the $(I)J^P=(0)0^-$ channel, the binding energy of the $\phi NN$ three-body system is smaller compared with scheme $\rm I$, since $\phi N$ is in the ${}^2S_{1/2}$ channel. Analogously, the binding energy of the $\phi NN$ three-body system in the $(I)J^P=(0)1^-$ channel also becomes smaller in this scheme. In the $(I)J^P=(0)2^-$ channel, the binding energy of the $\phi NN$ three-body system remains unchanged compared to scheme $\rm I$, since in this channel the subsystem $\phi N$ is in the ${}^4S_{3/2}$ channel.

\section{Conclusions}
\label{sec:conclusions}
In this work, Faddeev equations for the $J/\psi\,NN$ and $\phi\,NN$ three-body systems are constructed in momentum space. The subsystem two-body $t$-matrix is obtained by solving the Lippmann-Schwinger equations. The $NN$ interaction is constructed using the MT potential. The $J/\psi\,N$ interaction potential in the ${}^4S_{3/2}$ and ${}^2S_{1/2}$ channels is taken from HAL QCD. The $\phi\,N$ interaction potential in the ${}^4S_{3/2}$ channel is also taken from HAL QCD. For the $\phi N$ interaction in the ${}^2S_{1/2}$ channel, we have considered two schemes. In scheme $\rm I$, the $\phi N$ potential is obtained from the combination of the $\phi p$ correlation function analysis and the HAL QCD results. In scheme $\rm II$, the $\phi N$ potential in the ${}^2S_{1/2}$ channel is constructed under the assumption that the spin-spin parts of the $\phi N$ and $J/\psi N$ interactions are inversely proportional to the respective hadron masses.

Solving the Lippmann-Schwinger equations, we find that, with the experimental masses, the MT potential reproduces fairly well the deuteron energy. The scattering lengths obtained match the experimental values. When we use the masses employed in HAL QCD, the results are slightly modified. The $J/\psi N$ interaction in neither the ${}^4S_{3/2}$ nor the ${}^2S_{1/2}$ channel generates a bound state with the HAL QCD potentials. Using the masses employed in HAL QCD, the Lippmann-Schwinger equation reproduces the scattering length extracted by HAL QCD for the $J/\psi N$ interaction. The $\phi N$ interaction in the ${}^4S_{3/2}$ channel does not form a bound state with the HAL QCD potential. The scattering length and effective range obtained in this work in this channel agree with the HAL QCD results. In the ${}^2S_{1/2}$ channel, the $\phi N$ interaction in scheme $\rm I$ is strong enough to form a bound state. 

Solving the Faddeev equations, we did not find a bound state in the $J/\psi\,NN$ three-body system. For the $\phi\,NN$ three-body system, we found bound states in the $(I)J^P=(0)0^-$, $(0)1^-$, and $(0)2^-$ channels in both schemes $\rm I$ and $\rm II$. Compared to scheme $\rm I$, the binding energy of the $\phi NN$ three-body system is smaller in the $(I)J^P=(0)0^-$ and $(0)1^-$ channels in scheme $\rm II$, since the $\phi N$ interaction becomes weaker in this scheme. The binding energy of the $\phi NN$ three-body system remains unchanged in the $(I)J^P=(0)2^-$ channel, because the subsystem $\phi N$ is in the ${}^4S_{3/2}$ channel and remains unchanged. The predicted bound states in this work agree with the findings of Refs.~\cite{Wen:2025wit,Lazauskas:2026nlg,Lazauskas:2026ysy}. The possible bound states of the $\phi NN$ three-body system may be explored in photon-induced $\phi$ meson production on the deuteron in the reaction $\gamma d\to \phi d$ near threshold.

\begin{acknowledgments}
I wish to thank Feng-Kun Guo, Jia-Jun Wu and Bing-Song Zou for helpful discussions. This work is supported in part by the National Natural Science Foundation of China (NSFC) under Grants No. 12405106 and No. 12247139.
\end{acknowledgments}

\appendix
\section{Geometrical quantities}
\label{app:permuope}
\begin{widetext}
We use the Balian-Br\'ezin techniques~\cite{Balian:1969sd} to calculate the matrix element ${}_{1}\langle pq\alpha\mid p'q'\alpha'\rangle_{2}$ in Section~\ref{sec:pwbasis}. The details of the derivation with identical spin $1/2$ particles have already been given in Ref.~\cite{Gloeckle1983}. We evaluate the matrix element for particles with different masses,
\begin{equation}
X_{12}
\equiv {}_{1}\langle pq(l\lambda)LM \mid p'q'(l'\lambda')L'M'\rangle_{2}.
\end{equation}
Inserting two complete sets of three-momentum eigenstates, we can get
\begin{align}
X_{12}
&= \int d\bm{p}_1\,d\bm{q}_1\int d\bm{p}_1'\,d\bm{q}_1'\;
   {}_{1}\langle pq(l\lambda)LM\mid \bm{p}_1\bm{q}_1\rangle_{1}\;
   {}_{1}\langle \bm{p}_1\bm{q}_1\mid \bm{p}_1'\bm{q}_1'\rangle_{2}{}_{2}\langle \bm{p}_1'\bm{q}_1'\mid p'q'(l'\lambda')L'M'\rangle_{2}\nonumber\\
&= \int d\bm{p}_1\,d\bm{q}_1\int d\bm{p}_1'\,d\bm{q}_1'\;
   \mathcal{Y}_{l\lambda}^{LM*}(\hat{\bm{p}}_1\hat{\bm{q}}_1)\,
   \frac{\delta(p-p_1)}{p^2}\,\frac{\delta(q-q_1)}{q^2}\nonumber\\
&\quad\times
   {}_{1}\langle \bm{p}_1\bm{q}_1\mid  \bm{p}_1'\bm{q}_1'\rangle_{2}\;
   \mathcal{Y}_{l'\lambda'}^{L'M'}(\hat{\bm{p}}_1'\hat{\bm{q}}_1')\,
   \frac{\delta(p_1'-p')}{p'^2}\,\frac{\delta(q_1'-q')}{q'^2}.
\end{align}
Using the relation $\langle \bm{p}_1'\bm{q}_1'\mid \bm{p}_1\bm{q}_1\rangle = \delta^3(\bm{p}_1-\bm{p}_1')\delta^3(\bm{q}_1-\bm{q}_1')$, the overlap of the three-momentum eigenstates between different arrangements is
\begin{align}
{}_{1}\langle \bm{p}_1\bm{q}_1\mid \bm{p}_1'\bm{q}_1'\rangle_{2}
&= \delta^3\!\left(\bm{p}_1 + \frac{m_2}{m_{23}}\bm{p}_1'
                   - \frac{m_3 m_{123}}{m_{13}m_{23}}\bm{q}_1'\right)
   \delta^3\!\left(\bm{q}_1 + \bm{p}_1'
                   + \frac{m_1}{m_{13}}\bm{q}_1'\right)\nonumber \\
&= \delta^3\!\left(\bm{p}_1 - \frac{m_2}{m_{23}}\bm{q}_1
                   - \bm{q}_1'\right)
   \delta^3\!\left(\bm{p}_1'+\bm{q}_1 
                   + \frac{m_1}{m_{13}}\bm{q}_1'\right).                  
\end{align}

Then we can get 
\begin{align}
{X}_{12}
&= \int d{\bm{q}}_1\,d{\bm{q}}_1'\;
   \mathcal{Y}_{l\lambda}^{LM*}\!\left(\,
     \tfrac{m_2}{m_{23}}\widehat{\bm{q}_1+\bm{q}_1'},\,\hat{\bm{q}}_1\right)
   \frac{\delta(p-\pi_{12})}{p^2}\, \frac{\delta(q-q_1)}{q^2}\,\frac{\delta(p'-\pi_{12}')}{p'^2}\,\frac{\delta(q'-q_1')}{q^{'2}}\,
   \mathcal{Y}_{l'\lambda'}^{L'M'}\!\left(-\tfrac{m_1}{m_{13}}\widehat{\bm{q}_1'-\bm{q}_1},\,\hat{\bm{q}}_1'\right)\nonumber\\
&= \int d\hat{\bm{q}}\,d\hat{\bm{q}}'\;
   \mathcal{Y}_{l\lambda}^{LM*}\!\left(\,
     \tfrac{m_2}{m_{23}}\widehat{\bm{q}+\bm{q}'},\,\hat{\bm{q}}\right)
   \frac{\delta(p-\pi_{12})}{p^2}\,\frac{\delta(p'-\pi_{12}')}{p'^2}\,
   \mathcal{Y}_{l'\lambda'}^{L'M'}\!\left(-\tfrac{m_1}{m_{13}}\widehat{\bm{q}'-\bm{q}},\,\hat{\bm{q}}'\right).
\end{align}

The angular integral cannot be performed directly because the first argument of $\widehat{\bm{a}+\bm{b}} \equiv \frac{\bm{a}+\bm{b}}{|\bm{a}+\bm{b}|}$ appears in the spherical harmonics. Using the relation
\begin{equation}
Y_{lm}(\bm{a}+\bm{b})
= \sum_{l_1+l_2=l}
  \frac{a^{l_1}b^{l_2}}{|\bm{a}+\bm{b}|^l}
  \sqrt{\frac{4\pi(2l+1)!}{(2l_1+1)!\,(2l_2+1)!}}\;
  \mathcal{Y}_{l_1 l_2}^{lm}(\hat{\bm{a}}\hat{\bm{b}}),
\end{equation}
we obtain
\begin{align}
\mathcal{Y}_{l\lambda}^{LM}\!\left(\,
     \tfrac{m_2}{m_{23}}\widehat{\bm{q}+\bm{q}'},\,\hat{\bm{q}}\right)
&= \sum_{l_1+l_2=l}
   \frac{(\tfrac{m_2}{m_{23}}q)^{l_1}q'^{l_2}}{\pi_{12}^l}
   \sqrt{\frac{4\pi(2l+1)!}{(2l_1+1)!\,(2l_2+1)!}}\;
   \Bigl\{\mathcal{Y}_{l_1 l_2}^{l}(\hat{\bm{q}}\hat{\bm{q}}')\,Y_\lambda(\hat{\bm{q}})\Bigr\}^{LM}.
\end{align}
The expression in the bracket can be written as
\begin{align}
\Bigl\{\mathcal{Y}_{l_1 l_2}^{l}(\hat{\bm{q}}\hat{\bm{q}}')\,Y_\lambda(\hat{\bm{q}})\Bigr\}^{LM}
&= \sum_f (-1)^{l+\lambda-L}\sqrt{\hat{l}\hat{f}}\,
   \begin{Bmatrix}l_2 & l_1 & l \\ \lambda & L & f\end{Bmatrix}
   \Bigl\{Y_{l_2}(\hat{\bm{q}}')\,\mathcal{Y}_{l_1\lambda}^{f}(\hat{\bm{q}}\hat{\bm{q}})\Bigr\}^{LM}
\nonumber\\
&= \sum_f (-1)^{l+\lambda-L}\sqrt{\hat{l}\hat{l}_1\hat{\lambda}}\,
   \frac{1}{\sqrt{4\pi}}\,
   \begin{Bmatrix}l_2 & l_1 & l \\ \lambda & L & f\end{Bmatrix}
   C(l_1\lambda f,\,00)\;
   \mathcal{Y}_{l_2 f}^{LM}(\hat{\bm{q}}'\hat{\bm{q}}),
\end{align}
where we have used the relation
\begin{equation}
\mathcal{Y}_{l_1 l_2}^{lm}(\hat{\bm{a}}\hat{\bm{a}})
= \sqrt{\frac{\hat{l}_1\hat{l}_2}{4\pi\hat{l}}}\;C(l_1 l_2 l,\,00)\;Y_{lm}(\hat{\bm{a}}).
\end{equation}
Then we can get
\begin{align}
\mathcal{Y}_{l\lambda}^{LM}\!\left(\,
     \tfrac{m_2}{m_{23}}\widehat{\bm{q}+\bm{q}'},\,\hat{\bm{q}}\right)
&= \sum_{\substack{l_1+l_2=l}}
   \frac{(\tfrac{m_2}{m_{23}}q)^{l_1}q'^{l_2}}{\pi_{12}^l}
   \sqrt{\frac{(2l+1)!}{(2l_1+1)!\,(2l_2+1)!}}\;
   (-1)^{l+\lambda-L}\sqrt{\hat{l}\hat{l}_1\hat{\lambda}}\nonumber\\
&\quad\times
   \sum_f \begin{Bmatrix}l_2 & l_1 & l \\ \lambda & L & f\end{Bmatrix}
   C(l_1\lambda f,\,00)\;
   \mathcal{Y}_{l_2 f}^{LM}(\hat{\bm{q}}'\hat{\bm{q}}).
\end{align}

Analogously,
\begin{align}
\mathcal{Y}_{l'\lambda'}^{L'M'}\!\left(-\tfrac{m_1}{m_{13}}\widehat{\bm{q}'-\bm{q}},\,\hat{\bm{q}}'\right)
&= (-1)^{l'}
   \sum_{\substack{l_1'+l_2'=l'}}
   \frac{(\tfrac{m_1}{m_{13}}q')^{l_1'}q^{l_2'}}{\pi_{12}'^{l'}}
   \sqrt{\frac{(2l'+1)!}{(2l_1'+1)!\,(2l_2'+1)!}}\;
   \sqrt{\hat{l}'\hat{l}_1'\hat{\lambda}'}\nonumber\\
&\quad\times
   \sum_{f'} \begin{Bmatrix}l_2' & l_1' & l' \\ \lambda' & L' & f'\end{Bmatrix}
   C(l_1'\lambda'f',\,00)\;
   \mathcal{Y}_{f'l_2'}^{L'M'}(\hat{\bm{q}}'\hat{\bm{q}}).
\end{align}
The remaining angular dependence is due to the $\pi_{12}$ and $\pi_{12}'$.
We expand the combination as a Legendre series,
\begin{equation}
\frac{\delta(p-\pi_{12})}{p^{l+2}}\,\frac{\delta(p'-\pi_{12}')}{p'^{l'+2}}
= \sum_k 2\pi\sqrt{\hat{k}}\,(-1)^k\,g_k\;\mathcal{Y}_{kk}^{00}(\hat{\bm{q}}'\hat{\bm{q}}),
\end{equation}
where the coefficients $g_k$ are defined as
\begin{equation}
g_k = \int_{-1}^{1}dx\;P_k(x)\;
       \frac{\delta\!\left(p-\sqrt{\tfrac{m_2^2}{m_{23}^2}q^2+q'^2+\tfrac{2m_2}{m_{23}}qq'x}\right)}{p^{l+2}}\;
  \frac{\delta\!\left(p'-\sqrt{q^2+\tfrac{m_1^2}{m_{13}^2}q'^2+\tfrac{2m_1}{m_{13}}qq'x}\right)}{p'^{l'+2}}.
\end{equation}
Furthermore, we can reduce the product of two spherical harmonics to one,
\begin{align}
\mathcal{Y}_{kk}^{00}(\hat{\bm{q}}'\hat{\bm{q}})\;
\mathcal{Y}_{f'l_2'}^{L'M'}(\hat{\bm{q}}'\hat{\bm{q}})
&= \sum_{f_1 f_2}\sqrt{\hat{L}'\hat{f}_1\hat{f}_2}\,
   \,
   \begin{Bmatrix}k & k & 0 \\ f' & l_2' & L' \\ f_1 & f_2 & L'\end{Bmatrix} \Big\{\mathcal{Y}_{kf'}^{f_1}(\hat{\bm{q}}'\hat{\bm{q}}')\;
\mathcal{Y}_{kl_2'}^{f_2}(\hat{\bm{q}}\hat{\bm{q}})\Bigr\}^{L'M'}
\nonumber\\
&= \frac{1}{4\pi}\sqrt{\hat{k}\hat{f}'\hat{l}_2'}(-1)^{f'+l_2'+L'}
   \sum_{f_1 f_2}\begin{Bmatrix}f_2 & f_1 & L' \\ f' & l_2' & k\end{Bmatrix}
   C(kf'f_1,00)\,C(kl_2'f_2,00)\,\mathcal{Y}_{f_1 f_2}^{L'M'}(\hat{\bm{q}}'\hat{\bm{q}}).
\end{align}
Performing the angular integrations of $\hat{\bm{q}}$ and $\hat{\bm{q}}'$, we get
\begin{align}
&\int d\hat{\bm{q}}\,d\hat{\bm{q}}'\;
\mathcal{Y}_{l_2 f}^{LM*}(\hat{\bm{q}}'\hat{\bm{q}})\;
\mathcal{Y}_{kk}^{00}(\hat{\bm{q}}'\hat{\bm{q}})\;
\mathcal{Y}_{f'l_2'}^{L'M'}(\hat{\bm{q}}'\hat{\bm{q}}) =\frac{1}{4\pi}\sqrt{\hat{k}\hat{f}'\hat{l}_2'}\,
  (-1)^{f'+l_2'+L}
  \begin{Bmatrix}f & l_2 & L \\ f' & l_2' & k\end{Bmatrix}
  C(kf'l_2,00)\,C(kl_2'f,00)\,\delta_{LL'}\delta_{MM'}.
\end{align}
Collecting all ingredients, we have
\begin{align}
{}_{1}\langle pq(l\lambda)LM \mid p'q'(l'\lambda')L'M'\rangle_2
&= \delta_{LL'}\delta_{MM'}
   \sum_k g_k
   \sum_{\substack{l_1+l_2=l}}\, \sum_{\substack{l_1'+l_2'=l'}}
   q^{l_2+l_2'}\,q'^{l_1+l_1'}
   (-1)^{l'}\sqrt{\hat{l}\,\hat{\lambda}\,\hat{l}'\,\hat{\lambda}'\,}\,\hat{k}
   (\tfrac{m_2}{m_{23}})^{l_2}(\tfrac{m_1}{m_{13}})^{l_1'}(\tfrac{1}{2})\nonumber\\
&\quad\times
   \sqrt{\frac{(2l+1)!}{(2l_1)!\,(2l_2)!}}\;
   \sqrt{\frac{(2l'+1)!}{(2l_1')!\,(2l_2')!}}\;
   \sum_{ff'}\left\{\begin{matrix}l_1 & l_2 & l\\\lambda & L & f\end{matrix}\right\}C(l_2\lambda f,\,00)\;\nonumber\\
&\quad\times
   \left\{\begin{matrix}l_2' & l_1' & l'\\\lambda' & L & f'\end{matrix}\right\}
   C(l_1'\lambda'f',\,00)\left\{\begin{matrix}f & l_1 & L\\ f' & l_2' & k\end{matrix}\right\}\;C(kl_1f',\,00)\;C(kl_2'f,\,00).
\end{align}
In the $jI$-coupling, we can get
\begin{align}
{}_{1}\langle pq\alpha\mid p'q'\alpha'\rangle_{2}
&= \sum_{LS}\sum_{L'S'}
   \sqrt{\hat{j}\,\hat{I}\,\hat{L}\,\hat{S}}\;
   \sqrt{\hat{j}'\,\hat{I}'\,\hat{L}'\,\hat{S}'}\;
   \begin{Bmatrix}l & s & j \\ \lambda & s_1 & I \\ L & S & J\end{Bmatrix}
   \begin{Bmatrix}l' & s' & j' \\ \lambda' & s_2 & I' \\ L' & S' & J'\end{Bmatrix}
\nonumber\\
&\times\;
   {}_{1}\langle pq(l\lambda)LM\mid p'q'(l'\lambda')L'M'\rangle_{2}
  \,
   {}_{1}\langle (ss_1)S\mid (s's_2)S'\rangle_{2}
   \,
   {}_{1}\langle (tt_1)T\mid (t't_2)T'\rangle_{2}.
\end{align}
The spin and isospin matrix elements are recoupling coefficients between three-particles and can be evaluated as shown in Ref.~\cite{Edmonds1957}. Finally we can get  

\begin{align}
{}_{1}\langle pq\alpha\mid p'q'\alpha'\rangle_{2}
=\int_{-1}^1 dx  
       \frac{\delta\!\left(p-\sqrt{\tfrac{m_2^2}{m_{23}^2}q^2+q'^2+\tfrac{2m_2}{m_{23}}qq'x}\right)}{p^{l+2}}\;
  \frac{\delta\!\left(p'-\sqrt{q^2+\tfrac{m_1^2}{m_{13}^2}q'^2+\tfrac{2m_1}{m_{13}}qq'x}\right)}{p'^{l'+2}} G_{\alpha\alpha'}^{12}(q,q',x),
\end{align}
where geometrical quantities $G_{\alpha\alpha'}^{12}(q,q',x)$ is 
\begin{align}
G_{\alpha\alpha'}^{12}(q,q',x)=&\sum_{k}P_k(x) 
   \sum_{\substack{l_1+l_2=l}}\, \sum_{\substack{l_1'+l_2'=l'}}
   q^{l_2+l_2'}\,q'^{l_1+l_1'} \sqrt{\hat{l}\hat{s}\hat{j}\,\hat{t}\,\hat{\lambda}\,\hat{I}}\;
   \sqrt{\hat{l}'\hat{s}'\hat{j}'\,\hat{t}'\,\hat{\lambda}'\,\hat{I}'}\; \left\{\begin{matrix} t_2 & t_3 & t\\ t_1 & T & t'\end{matrix}\right\} \nonumber \\
 & \times  \,(-1)^{s_1+2s_2+s_3}\,(-1)^{t_1+2t_2+t_3} \, (-1)^{s'+l'+t'}\,\hat{k}
   (\tfrac{m_2}{m_{23}})^{l_2}(\tfrac{m_1}{m_{13}})^{l_1'}(\tfrac{1}{2}) \nonumber \\
&\times\sum_{LS} \hat{L} \hat{S} \; \left\{\begin{matrix} s_2 & s_3 & s\\ s_1 & S & s'\end{matrix}\right\}
   \begin{Bmatrix}l & s & j \\ \lambda & s_1 & I \\ L & S & J\end{Bmatrix}
   \begin{Bmatrix}l' & s' & j' \\ \lambda' & s_2 & I' \\ L & S & J\end{Bmatrix}
  \nonumber\\
&\quad\times
   \sqrt{\frac{(2l+1)!}{(2l_1)!\,(2l_2)!}}\;
   \sqrt{\frac{(2l'+1)!}{(2l_1')!\,(2l_2')!}}\;
   \sum_{ff'}\left\{\begin{matrix}l_1 & l_2 & l\\\lambda & L & f\end{matrix}\right\}C(l_2\lambda f,\,00)\;\nonumber\\
&\quad\times
   \left\{\begin{matrix}l_2' & l_1' & l'\\\lambda' & L & f'\end{matrix}\right\}
   C(l_1'\lambda'f',\,00)\left\{\begin{matrix}f & l_1 & L\\ f' & l_2' & k\end{matrix}\right\}\;C(kl_1f',\,00)\;C(kl_2'f,\,00).
\end{align}
\end{widetext}

\bibliography{ref.bib}

@article{Gross:1973id,
    author = "Gross, David J. and Wilczek, Frank",
    title = "{Ultraviolet Behavior of Nonabelian Gauge Theories}",
    doi = "10.1103/PhysRevLett.30.1343",
    journal = "Phys. Rev. Lett.",
    volume = "30",
    pages = "1343--1346",
    year = "1973"
}

@article{Politzer:1973fx,
    author = "Politzer, H. David",
    title = "{Reliable Perturbative Results for Strong Interactions?}",
    doi = "10.1103/PhysRevLett.30.1346",
    journal = "Phys. Rev. Lett.",
    volume = "30",
    pages = "1346--1349",
    year = "1973"
}

@article{Gross:2022hyw,
    author = "Gross, Franz and others",
    title = "{50 Years of Quantum Chromodynamics}",
    eprint = "2212.11107",
    archivePrefix = "arXiv",
    primaryClass = "hep-ph",
    doi = "10.1140/epjc/s10052-023-11949-2",
    journal = "Eur. Phys. J. C",
    volume = "83",
    pages = "1125",
    year = "2023"
}

@article{Kharzeev:1995ij,
    author = "Kharzeev, D.",
    editor = "Di Giacomo, A. and Diakonov, Dmitri",
    title = "{Quarkonium interactions in QCD}",
    eprint = "nucl-th/9601029",
    archivePrefix = "arXiv",
    reportNumber = "CERN-TH-95-342, BI-TP-95-41",
    doi = "10.3254/978-1-61499-215-8-105",
    journal = "Proc. Int. Sch. Phys. Fermi",
    volume = "130",
    pages = "105--131",
    year = "1996"
}

@article{Kharzeev:1998bz,
    author = "Kharzeev, D. and Satz, H. and Syamtomov, A. and Zinovjev, G.",
    title = "{J / psi photoproduction and the gluon structure of the nucleon}",
    eprint = "hep-ph/9901375",
    archivePrefix = "arXiv",
    reportNumber = "BI-TP-98-36",
    doi = "10.1007/s100529900047",
    journal = "Eur. Phys. J. C",
    volume = "9",
    pages = "459--462",
    year = "1999"
}

@article{Gittelman:1975ix,
    author = "Gittelman, B. and Hanson, K. M. and Larson, D. and Loh, E. and Silverman, A. and Theodosiou, G.",
    title = "{Photoproduction of the psi (3100) Meson at 11-GeV}",
    reportNumber = "CLNS 312",
    doi = "10.1103/PhysRevLett.35.1616",
    journal = "Phys. Rev. Lett.",
    volume = "35",
    pages = "1616",
    year = "1975"
}

@article{Camerini:1975cy,
    author = "Camerini, U. and Learned, J. G. and Prepost, R. and Spencer, Cherrill M. and Wiser, D. E. and Ash, W. and Anderson, Robert L. and Ritson, D. and Sherden, D. and Sinclair, Charles K.",
    title = "{Photoproduction of the psi Particles}",
    reportNumber = "SLAC-PUB-1591, COO-881-461",
    doi = "10.1103/PhysRevLett.35.483",
    journal = "Phys. Rev. Lett.",
    volume = "35",
    pages = "483",
    year = "1975"
}

@article{GlueX:2019mkq,
    author = "Ali, A. and others",
    collaboration = "GlueX",
    title = "{First Measurement of Near-Threshold J/{\ensuremath{\psi}} Exclusive Photoproduction off the Proton}",
    eprint = "1905.10811",
    archivePrefix = "arXiv",
    primaryClass = "nucl-ex",
    reportNumber = "JLAB-PHY-19-2946",
    doi = "10.1103/PhysRevLett.123.072001",
    journal = "Phys. Rev. Lett.",
    volume = "123",
    number = "7",
    pages = "072001",
    year = "2019"
}

@article{GlueX:2023pev,
    author = "Adhikari, S. and others",
    collaboration = "GlueX",
    title = "{Measurement of the J/$\psi $ photoproduction cross section over the full near-threshold kinematic region}",
    eprint = "2304.03845",
    archivePrefix = "arXiv",
    primaryClass = "nucl-ex",
    reportNumber = "JLAB-PHY-23-3897",
    doi = "10.1103/PhysRevC.108.025201",
    journal = "Phys. Rev. C",
    volume = "108",
    number = "2",
    pages = "025201",
    year = "2023"
}

@article{Duran:2022xag,
    author = "Duran, B. and others",
    title = "{Determining the gluonic gravitational form factors of the proton}",
    eprint = "2207.05212",
    archivePrefix = "arXiv",
    primaryClass = "nucl-ex",
    doi = "10.1038/s41586-023-05730-4",
    journal = "Nature",
    volume = "615",
    number = "7954",
    pages = "813--816",
    year = "2023"
}

@article{007:2026dow,
    author = "Joosten, S. and others",
    collaboration = "007",
    title = "{Near-Threshold J/$ψ\to μ^+μ^-$ Photoproduction and the Gluonic Gravitational Form Factors of the Proton}",
    eprint = "2602.14416",
    archivePrefix = "arXiv",
    primaryClass = "nucl-ex",
    month = "2",
    year = "2026"
}

@article{CLAS:2026lls,
    author = "Chatagnon, P. and others",
    collaboration = "CLAS",
    title = "{Measurement of the near-threshold J/{\ensuremath{\psi}} photoproduction cross section with the CLAS12 experiment}",
    eprint = "2602.22128",
    archivePrefix = "arXiv",
    primaryClass = "hep-ex",
    reportNumber = "JLAB-PHY-26-4605",
    doi = "10.1103/lqv1-kf2n",
    journal = "Phys. Rev. C",
    volume = "113",
    number = "6",
    pages = "065203",
    year = "2026"
}

@article{Gell-Mann:1961jim,
    author = "Gell-Mann, Murray and Zachariasen, Fredrik",
    title = "{Form-factors and vector mesons}",
    doi = "10.1103/PhysRev.124.953",
    journal = "Phys. Rev.",
    volume = "124",
    pages = "953--964",
    year = "1961"
}

@article{Kroll:1967it,
    author = "Kroll, N. M. and Lee, T. D. and Zumino, B.",
    editor = "Feinberg, G.",
    title = "{Neutral Vector Mesons and the Hadronic Electromagnetic Current}",
    doi = "10.1103/PhysRev.157.1376",
    journal = "Phys. Rev.",
    volume = "157",
    pages = "1376--1399",
    year = "1967"
}

@article{Kuraev:1977fs,
    author = "Kuraev, E. A. and Lipatov, L. N. and Fadin, Victor S.",
    title = "{The Pomeranchuk Singularity in Nonabelian Gauge Theories}",
    journal = "Sov. Phys. JETP",
    volume = "45",
    pages = "199--204",
    year = "1977"
}

@article{Balitsky:1978ic,
    author = "Balitsky, I. I. and Lipatov, L. N.",
    title = "{The Pomeranchuk Singularity in Quantum Chromodynamics}",
    journal = "Sov. J. Nucl. Phys.",
    volume = "28",
    pages = "822--829",
    year = "1978"
}

@article{Bauer:1977iq,
    author = "Bauer, T. H. and Spital, R. D. and Yennie, D. R. and Pipkin, F. M.",
    title = "{The Hadronic Properties of the Photon in High-Energy Interactions}",
    reportNumber = "PRINT-77-0549 (HARVARD)",
    doi = "10.1103/RevModPhys.50.261",
    journal = "Rev. Mod. Phys.",
    volume = "50",
    pages = "261",
    year = "1978",
    note = "[Erratum: Rev.Mod.Phys. 51, 407 (1979)]"
}

@article{Pumplin:1975fd,
    author = "Pumplin, Jon and Repko, Wayne",
    title = "{Diffractive Production of the psi Particle and Charmed Mesons in Photon and Neutrino Interactions}",
    reportNumber = "Print-75-0335 (MICHIGAN STATE)",
    doi = "10.1103/PhysRevD.12.1376",
    journal = "Phys. Rev. D",
    volume = "12",
    pages = "1376",
    year = "1975"
}

@article{Barger:1975ng,
    author = "Barger, Vernon D. and Phillips, R. J. N.",
    title = "{Properties of psi n Scattering}",
    reportNumber = "COO-881-462",
    doi = "10.1016/0370-2693(75)90582-1",
    journal = "Phys. Lett. B",
    volume = "58",
    pages = "433--436",
    year = "1975"
}

@article{Peskin:1979va,
    author = "Peskin, Michael E.",
    title = "{Short Distance Analysis for Heavy Quark Systems. 1. Diagrammatics}",
    reportNumber = "HUTP-79/A008",
    doi = "10.1016/0550-3213(79)90199-8",
    journal = "Nucl. Phys. B",
    volume = "156",
    pages = "365--390",
    year = "1979"
}

@article{Bhanot:1979vb,
    author = "Bhanot, Gyan and Peskin, Michael E.",
    title = "{Short Distance Analysis for Heavy Quark Systems. 2. Applications}",
    reportNumber = "HUTP-79/A015",
    doi = "10.1016/0550-3213(79)90200-1",
    journal = "Nucl. Phys. B",
    volume = "156",
    pages = "391--416",
    year = "1979"
}

@article{Voloshin:1978hc,
    author = "Voloshin, M. B.",
    title = "{On Dynamics of Heavy Quarks in Nonperturbative QCD Vacuum}",
    reportNumber = "ITEP-86-1978",
    doi = "10.1016/0550-3213(79)90037-3",
    journal = "Nucl. Phys. B",
    volume = "154",
    pages = "365--380",
    year = "1979"
}

@article{Gottfried:1977gp,
    author = "Gottfried, Kurt",
    title = "{Hadronic Transitions Between Quark - anti-Quark Bound States}",
    reportNumber = "CLNS-381",
    doi = "10.1103/PhysRevLett.40.598",
    journal = "Phys. Rev. Lett.",
    volume = "40",
    pages = "598",
    year = "1978"
}

@article{Appelquist:1978rt,
    author = "Appelquist, Thomas and Fischler, W.",
    title = "{Some Remarks on Van Der Waals Forces in {QCD}}",
    reportNumber = "Print-78-0351 (YALE)",
    doi = "10.1016/0370-2693(78)90587-7",
    journal = "Phys. Lett. B",
    volume = "77",
    pages = "405--410",
    year = "1978"
}

@article{Luke:1992tm,
    author = "Luke, Michael E. and Manohar, Aneesh V. and Savage, Martin J.",
    title = "{A QCD Calculation of the interaction of quarkonium with nuclei}",
    eprint = "hep-ph/9204219",
    archivePrefix = "arXiv",
    reportNumber = "UCSD-PTH-92-12",
    doi = "10.1016/0370-2693(92)91114-O",
    journal = "Phys. Lett. B",
    volume = "288",
    pages = "355--359",
    year = "1992"
}

@article{Ji:1994av,
    author = "Ji, Xiang-Dong",
    title = "{A QCD analysis of the mass structure of the nucleon}",
    eprint = "hep-ph/9410274",
    archivePrefix = "arXiv",
    reportNumber = "MIT-CTP-2368",
    doi = "10.1103/PhysRevLett.74.1071",
    journal = "Phys. Rev. Lett.",
    volume = "74",
    pages = "1071--1074",
    year = "1995"
}

@article{Hatta:2018ina,
    author = "Hatta, Yoshitaka and Yang, Di-Lun",
    title = "{Holographic $J/\psi$ production near threshold and the proton mass problem}",
    eprint = "1808.02163",
    archivePrefix = "arXiv",
    primaryClass = "hep-ph",
    reportNumber = "YITP-18-79",
    doi = "10.1103/PhysRevD.98.074003",
    journal = "Phys. Rev. D",
    volume = "98",
    number = "7",
    pages = "074003",
    year = "2018"
}

@article{Kou:2023zko,
    author = "Kou, Wei and Chen, Xurong",
    title = "{Unraveling proton strangeness: Determination of the strangeness sigma term with statistical significance}",
    eprint = "2311.08789",
    archivePrefix = "arXiv",
    primaryClass = "hep-ph",
    doi = "10.1103/PhysRevD.109.036034",
    journal = "Phys. Rev. D",
    volume = "109",
    number = "3",
    pages = "036034",
    year = "2024"
}

@article{Brodsky:1989jd,
    author = "Brodsky, Stanley J. and Schmidt, I. A. and de Teramond, G. F.",
    title = "{NUCLEAR BOUND QUARKONIUM}",
    reportNumber = "SLAC-PUB-5102",
    doi = "10.1103/PhysRevLett.64.1011",
    journal = "Phys. Rev. Lett.",
    volume = "64",
    pages = "1011",
    year = "1990"
}

@article{TarrusCastella:2018php,
    author = "Tarr{\'u}s Castell{\`a}, Jaume and Krein, Gast{\~a}o",
    title = "{Effective field theory for the nucleon-quarkonium interaction}",
    eprint = "1803.05412",
    archivePrefix = "arXiv",
    primaryClass = "hep-ph",
    reportNumber = "TUM-EFT 108/18, TUM-EFT-108-18",
    doi = "10.1103/PhysRevD.98.014029",
    journal = "Phys. Rev. D",
    volume = "98",
    number = "1",
    pages = "014029",
    year = "2018"
}

@article{Strakovsky:2019bev,
    author = "Strakovsky, Igor and Epifanov, Denis and Pentchev, Lubomir",
    title = "{J/$\psi$p scattering length from GlueX threshold measurements}",
    eprint = "1911.12686",
    archivePrefix = "arXiv",
    primaryClass = "hep-ph",
    doi = "10.1103/PhysRevC.101.042201",
    journal = "Phys. Rev. C",
    volume = "101",
    number = "4",
    pages = "042201",
    year = "2020"
}

@article{JointPhysicsAnalysisCenter:2023qgg,
    author = "Winney, D. and others",
    collaboration = "Joint Physics Analysis Center",
    title = "{Dynamics in near-threshold J/{\ensuremath{\psi}} photoproduction}",
    eprint = "2305.01449",
    archivePrefix = "arXiv",
    primaryClass = "hep-ph",
    reportNumber = "JLAB-THY-23-3802",
    doi = "10.1103/PhysRevD.108.054018",
    journal = "Phys. Rev. D",
    volume = "108",
    number = "5",
    pages = "054018",
    year = "2023"
}

@article{Wu:2024xwy,
    author = "Wu, Bing and Dong, Xiang-Kun and Du, Meng-Lin and Guo, Feng-Kun and Zou, Bing-Song",
    title = "{Deciphering the mechanism of J/{\ensuremath{\psi}}-nucleon scattering}",
    eprint = "2410.19526",
    archivePrefix = "arXiv",
    primaryClass = "hep-ph",
    doi = "10.1016/j.fmre.2025.07.005",
    journal = "Fund. Res.",
    volume = "5",
    pages = "2530--2536",
    year = "2025"
}

@article{LHCb:2015yax,
    author = "Aaij, Roel and others",
    collaboration = "LHCb",
    title = "{Observation of $J/\psi p$ Resonances Consistent with Pentaquark States in $\Lambda_b^0 \to J/\psi K^- p$ Decays}",
    eprint = "1507.03414",
    archivePrefix = "arXiv",
    primaryClass = "hep-ex",
    reportNumber = "CERN-PH-EP-2015-153, LHCB-PAPER-2015-029",
    doi = "10.1103/PhysRevLett.115.072001",
    journal = "Phys. Rev. Lett.",
    volume = "115",
    pages = "072001",
    year = "2015"
}

@article{LHCb:2016ztz,
    author = "Aaij, Roel and others",
    collaboration = "LHCb",
    title = "{Model-independent evidence for $J/\psi p$ contributions to $\Lambda_b^0\to J/\psi p K^-$ decays}",
    eprint = "1604.05708",
    archivePrefix = "arXiv",
    primaryClass = "hep-ex",
    reportNumber = "LHCB-PAPER-2016-009, CERN-EP-2016-086",
    doi = "10.1103/PhysRevLett.117.082002",
    journal = "Phys. Rev. Lett.",
    volume = "117",
    number = "8",
    pages = "082002",
    year = "2016"
}

@article{LHCb:2019kea,
    author = "Aaij, Roel and others",
    collaboration = "LHCb",
    title = "{Observation of a narrow pentaquark state, $P_c(4312)^+$, and of two-peak structure of the $P_c(4450)^+$}",
    eprint = "1904.03947",
    archivePrefix = "arXiv",
    primaryClass = "hep-ex",
    reportNumber = "LHCb-PAPER-2019-014 CERN-EP-2019-058",
    doi = "10.1103/PhysRevLett.122.222001",
    journal = "Phys. Rev. Lett.",
    volume = "122",
    number = "22",
    pages = "222001",
    year = "2019"
}

@article{Yokokawa:2006td,
    author = "Yokokawa, Kazuo and Sasaki, Shoichi and Hatsuda, Tetsuo and Hayashigaki, Arata",
    title = "{First lattice study of low-energy charmonium-hadron interaction}",
    eprint = "hep-lat/0605009",
    archivePrefix = "arXiv",
    doi = "10.1103/PhysRevD.74.034504",
    journal = "Phys. Rev. D",
    volume = "74",
    pages = "034504",
    year = "2006"
}

@article{Skerbis:2018lew,
    author = "Skerbis, Ursa and Prelovsek, Sasa",
    title = "{Nucleon-$J/\psi$ and nucleon-$\eta_{c}$ scattering in $P_{c}$ pentaquark channels from LQCD}",
    eprint = "1811.02285",
    archivePrefix = "arXiv",
    primaryClass = "hep-lat",
    doi = "10.1103/PhysRevD.99.094505",
    journal = "Phys. Rev. D",
    volume = "99",
    number = "9",
    pages = "094505",
    year = "2019"
}

@article{Lyu:2024ttm,
    author = "Lyu, Yan and Doi, Takumi and Hatsuda, Tetsuo and Sugiura, Takuya",
    title = "{Nucleon-charmonium interactions from lattice QCD}",
    eprint = "2410.22755",
    archivePrefix = "arXiv",
    primaryClass = "hep-lat",
    reportNumber = "RIKEN-iTHEMS-Report-24",
    doi = "10.1016/j.physletb.2024.139178",
    journal = "Phys. Lett. B",
    volume = "860",
    pages = "139178",
    year = "2025"
}

@book{Gloeckle1983,
    author    = {Glöckle, Walter},
    title     = {The Quantum Mechanical Few-Body Problem},
    publisher = {Springer},
    address   = {Berlin, Heidelberg},
    year      = {1983},
    doi       = {10.1007/978-3-642-82081-6},
}

@article{Gao:2000az,
    author = "Gao, H. and Lee, T. S. H. and Marinov, V.",
    title = "{Phi0 - N bound state}",
    eprint = "nucl-th/0010042",
    archivePrefix = "arXiv",
    doi = "10.1103/PhysRevC.63.022201",
    journal = "Phys. Rev. C",
    volume = "63",
    pages = "022201",
    year = "2001"
}

@article{Lyu:2022imf,
    author = "Lyu, Yan and Doi, Takumi and Hatsuda, Tetsuo and Ikeda, Yoichi and Meng, Jie and Sasaki, Kenji and Sugiura, Takuya",
    title = "{Attractive N-{\ensuremath{\phi}} interaction and two-pion tail from lattice QCD near physical point}",
    eprint = "2205.10544",
    archivePrefix = "arXiv",
    primaryClass = "hep-lat",
    reportNumber = "RIKEN-iTHEMS-Report-22",
    doi = "10.1103/PhysRevD.106.074507",
    journal = "Phys. Rev. D",
    volume = "106",
    number = "7",
    pages = "074507",
    year = "2022"
}

@article{Wen:2025wit,
    author = "Wen, Liang-Zhen and Ma, Yao and Meng, Lu and Zhu, Shi-Lin",
    title = "{{\ensuremath{\phi}}NN, J/{\ensuremath{\psi}}NN, {\ensuremath{\eta}}cNN systems based on HAL QCD interactions}",
    eprint = "2503.11938",
    archivePrefix = "arXiv",
    primaryClass = "hep-ph",
    doi = "10.1103/mvqk-n377",
    journal = "Phys. Rev. D",
    volume = "111",
    number = "11",
    pages = "114004",
    year = "2025",
    note = "[Erratum: Phys.Rev.D 112, 039901 (2025)]"
}

@article{Chizzali:2022pjd,
    author = "Chizzali, Emma and Kamiya, Yuki and Del Grande, Raffaele and Doi, Takumi and Fabbietti, Laura and Hatsuda, Tetsuo and Lyu, Yan",
    title = "{Indication of a p{\textendash}{\ensuremath{\phi}} bound state from a correlation function analysis}",
    eprint = "2212.12690",
    archivePrefix = "arXiv",
    primaryClass = "nucl-ex",
    reportNumber = "IPM/P-2012/009, RIKEN-iTHEMS-Report-22, MPP-2022-290",
    doi = "10.1016/j.physletb.2023.138358",
    journal = "Phys. Lett. B",
    volume = "848",
    pages = "138358",
    year = "2024"
}

@article{Belyaev:2007yc,
    author = "Belyaev, Vladimir B. and Sandhas, Werner and Shlyk, Ivan I.",
    editor = "Kievsky, Alejandro and Viviani, Michele",
    title = "{New nuclear three-body clusters {\textbackslash}phi{NN}}",
    eprint = "0707.4615",
    archivePrefix = "arXiv",
    primaryClass = "nucl-th",
    doi = "10.1007/s00601-008-0324-5",
    journal = "Few Body Syst.",
    volume = "44",
    pages = "347--349",
    year = "2008"
}

@article{Belyaev:2009ag,
    author = "Belyaev, V. B. and Sandhas, W. and Shlyk, I. I.",
    title = "{3- and 4- body meson- nuclear clusters}",
    eprint = "0903.1703",
    archivePrefix = "arXiv",
    primaryClass = "nucl-th",
    month = "3",
    year = "2009"
}

@article{Sofianos_2010,
doi = {10.1088/0954-3899/37/8/085109},
url = {https://doi.org/10.1088/0954-3899/37/8/085109},
year = {2010},
month = {jun},
publisher = {},
volume = {37},
number = {8},
pages = {085109},
author = {Sofianos, S A and Rampho, G J and Braun, M and Adam, R M},
title = {The ϕ–NN and ϕϕ–NN mesic nuclear systems},
journal = {Journal of Physics G: Nuclear and Particle Physics},
}

@article{Etminan:2024vkv,
    author = "Etminan, Faisal and Aalimi, Amanullah",
    title = "{Examination of the {\ensuremath{\phi}}-NN bound-state problem with lattice QCD N-{\ensuremath{\phi}} potentials}",
    eprint = "2402.06914",
    archivePrefix = "arXiv",
    primaryClass = "nucl-th",
    reportNumber = "RIKEN-iTHEMS-Report-23",
    doi = "10.1103/PhysRevC.109.054002",
    journal = "Phys. Rev. C",
    volume = "109",
    number = "5",
    pages = "054002",
    year = "2024"
}

@article{Filikhin:2024avj,
    author = "Filikhin, Igor and Kezerashvili, Roman Ya. and Vlahovic, Branislav",
    title = "{Possible H{\ensuremath{\phi}}3 hypernucleus with the HAL QCD interaction}",
    eprint = "2407.12190",
    archivePrefix = "arXiv",
    primaryClass = "nucl-th",
    doi = "10.1103/PhysRevD.110.L031502",
    journal = "Phys. Rev. D",
    volume = "110",
    number = "3",
    pages = "L031502",
    year = "2024"
}

@article{Faddeev:1960su,
    author = "Faddeev, L. D.",
    title = "{Scattering Theory for a Three-Particle System}",
    journal = "Sov. Phys. JETP",
    volume = "12",
    pages = "1014--1019",
    year = "1961"
}

@book{faddeev:1965ml,
    author = "Faddeev, L. D.",
    title ="Mathematical aspects of the three-body problem in the 
    quantum scattering theory",
    publisher ="Israel Program for Scientific Translations, Jerusalem",
    year ="1965" 
}

@article{Zhang:2024dkm,
    author = "Zhang, Xu",
    title = "{Pentaquarks $P_c$ in a dynamical coupled-channel approach of $\gamma p \rightarrow J/\psi p$ reaction}",
    eprint = "2410.10154",
    archivePrefix = "arXiv",
    primaryClass = "hep-ph",
    doi = "10.1140/epjc/s10052-025-14845-z",
    journal = "Eur. Phys. J. C",
    volume = "85",
    number = "10",
    pages = "1120",
    year = "2025"
}

@article{Wu:2010jy,
    author = "Wu, Jia-Jun and Molina, R. and Oset, E. and Zou, B. S.",
    title = "{Prediction of narrow $N^*$ and $\Lambda^*$ resonances with hidden charm above 4 GeV}",
    eprint = "1007.0573",
    archivePrefix = "arXiv",
    primaryClass = "nucl-th",
    doi = "10.1103/PhysRevLett.105.232001",
    journal = "Phys. Rev. Lett.",
    volume = "105",
    pages = "232001",
    year = "2010"
}

@article{Liu:2019tjn,
    author = "Liu, Ming-Zhu and Pan, Ya-Wen and Peng, Fang-Zheng and S\'anchez S\'anchez, Mario and Geng, Li-Sheng and Hosaka, Atsushi and Pavon Valderrama, Manuel",
    title = "{Emergence of a complete heavy-quark spin symmetry multiplet: seven molecular pentaquarks in light of the latest LHCb analysis}",
    eprint = "1903.11560",
    archivePrefix = "arXiv",
    primaryClass = "hep-ph",
    doi = "10.1103/PhysRevLett.122.242001",
    journal = "Phys. Rev. Lett.",
    volume = "122",
    number = "24",
    pages = "242001",
    year = "2019"
}

@article{Fernandez-Ramirez:2019koa,
    author = "Fern\'andez-Ram\'\i{}rez, C. and Pilloni, A. and Albaladejo, M. and Jackura, A. and Mathieu, V. and Mikhasenko, M. and Silva-Castro, J. A. and Szczepaniak, A. P.",
    collaboration = "JPAC",
    title = "{Interpretation of the LHCb $P_c$(4312)$^+$ Signal}",
    eprint = "1904.10021",
    archivePrefix = "arXiv",
    primaryClass = "hep-ph",
    reportNumber = "JLAB-THY-19-2921",
    doi = "10.1103/PhysRevLett.123.092001",
    journal = "Phys. Rev. Lett.",
    volume = "123",
    number = "9",
    pages = "092001",
    year = "2019"
}

@article{Du:2019pij,
    author = "Du, Meng-Lin and Baru, Vadim and Guo, Feng-Kun and Hanhart, Christoph and Mei\ss{}ner, Ulf-G and Oller, Jos\'e A. and Wang, Qian",
    title = "{Interpretation of the LHCb $P_c$ States as Hadronic Molecules and Hints of a Narrow $P_c(4380)$}",
    eprint = "1910.11846",
    archivePrefix = "arXiv",
    primaryClass = "hep-ph",
    doi = "10.1103/PhysRevLett.124.072001",
    journal = "Phys. Rev. Lett.",
    volume = "124",
    number = "7",
    pages = "072001",
    year = "2020"
}

@article{Chen:2019bip,
    author = "Chen, Hua-Xing and Chen, Wei and Zhu, Shi-Lin",
    title = "{Possible interpretations of the $P_c(4312)$, $P_c(4440)$, and $P_c(4457)$}",
    eprint = "1903.11001",
    archivePrefix = "arXiv",
    primaryClass = "hep-ph",
    doi = "10.1103/PhysRevD.100.051501",
    journal = "Phys. Rev. D",
    volume = "100",
    number = "5",
    pages = "051501",
    year = "2019"
}

@article{Chen:2019asm,
    author = "Chen, Rui and Sun, Zhi-Feng and Liu, Xiang and Zhu, Shi-Lin",
    title = "{Strong LHCb evidence supporting the existence of the hidden-charm molecular pentaquarks}",
    eprint = "1903.11013",
    archivePrefix = "arXiv",
    primaryClass = "hep-ph",
    doi = "10.1103/PhysRevD.100.011502",
    journal = "Phys. Rev. D",
    volume = "100",
    number = "1",
    pages = "011502",
    year = "2019"
}

@article{He:2019ify,
    author = "He, Jun",
    title = "{Study of $P_c(4457)$, $P_c(4440)$, and $P_c(4312)$ in a quasipotential Bethe-Salpeter equation approach}",
    eprint = "1903.11872",
    archivePrefix = "arXiv",
    primaryClass = "hep-ph",
    doi = "10.1140/epjc/s10052-019-6906-1",
    journal = "Eur. Phys. J. C",
    volume = "79",
    number = "5",
    pages = "393",
    year = "2019"
}

@article{Guo:2019kdc,
    author = "Guo, Zhi-Hui and Oller, J. A.",
    title = "{Anatomy of the newly observed hidden-charm pentaquark states: $P_c(4312)$, $P_c(4440)$ and $P_c(4457)$}",
    eprint = "1904.00851",
    archivePrefix = "arXiv",
    primaryClass = "hep-ph",
    doi = "10.1016/j.physletb.2019.04.053",
    journal = "Phys. Lett. B",
    volume = "793",
    pages = "144--149",
    year = "2019"
}

@article{Xiao:2019mvs,
    author = "Xiao, Cheng-Jian and Huang, Yin and Dong, Yu-Bing and Geng, Li-Sheng and Chen, Dian-Yong",
    title = "{Exploring the molecular scenario of Pc(4312) , Pc(4440) , and Pc(4457)}",
    eprint = "1904.00872",
    archivePrefix = "arXiv",
    primaryClass = "hep-ph",
    doi = "10.1103/PhysRevD.100.014022",
    journal = "Phys. Rev. D",
    volume = "100",
    number = "1",
    pages = "014022",
    year = "2019"
}

@article{Xiao:2019aya,
    author = "Xiao, C. W. and Nieves, J. and Oset, E.",
    title = "{Heavy quark spin symmetric molecular states from ${\bar D}^{(*)}\Sigma_c^{(*)}$ and other coupled channels in the light of the recent LHCb pentaquarks}",
    eprint = "1904.01296",
    archivePrefix = "arXiv",
    primaryClass = "hep-ph",
    doi = "10.1103/PhysRevD.100.014021",
    journal = "Phys. Rev. D",
    volume = "100",
    number = "1",
    pages = "014021",
    year = "2019"
}

@article{Wang:2019ato,
    author = "Wang, Bo and Meng, Lu and Zhu, Shi-Lin",
    title = "{Hidden-charm and hidden-bottom molecular pentaquarks in chiral effective field theory}",
    eprint = "1909.13054",
    archivePrefix = "arXiv",
    primaryClass = "hep-ph",
    doi = "10.1007/JHEP11(2019)108",
    journal = "JHEP",
    volume = "11",
    pages = "108",
    year = "2019"
}

@article{Meng:2019ilv,
    author = "Meng, Lu and Wang, Bo and Wang, Guang-Juan and Zhu, Shi-Lin",
    title = "{The hidden charm pentaquark states and $\Sigma_c\bar{D}^{(*)}$ interaction in chiral perturbation theory}",
    eprint = "1905.04113",
    archivePrefix = "arXiv",
    primaryClass = "hep-ph",
    doi = "10.1103/PhysRevD.100.014031",
    journal = "Phys. Rev. D",
    volume = "100",
    number = "1",
    pages = "014031",
    year = "2019"
}

@article{Voloshin:2019aut,
    author = "Voloshin, M. B.",
    title = "{Some decay properties of hidden-charm pentaquarks as baryon-meson molecules}",
    eprint = "1907.01476",
    archivePrefix = "arXiv",
    primaryClass = "hep-ph",
    reportNumber = "FTPI-MINN-19/19, UMN-TH-3828/19",
    doi = "10.1103/PhysRevD.100.034020",
    journal = "Phys. Rev. D",
    volume = "100",
    number = "3",
    pages = "034020",
    year = "2019"
}

@article{Wang:2019hyc,
    author = "Wang, Zhi-Gang and Wang, Xu",
    title = "{Analysis of the strong decays of the $P_c(4312)$ as a pentaquark molecular state with QCD sum rules}",
    eprint = "1907.04582",
    archivePrefix = "arXiv",
    primaryClass = "hep-ph",
    doi = "10.1088/1674-1137/ababf7",
    journal = "Chin. Phys. C",
    volume = "44",
    pages = "103102",
    year = "2020"
}

@article{Yamaguchi:2019seo,
    author = "Yamaguchi, Yasuhiro and Garc\'\i{}a-Tecocoatzi, Hugo and Giachino, Alessandro and Hosaka, Atsushi and Santopinto, Elena and Takeuchi, Sachiko and Takizawa, Makoto",
    title = "{$P_c$ pentaquarks with chiral tensor and quark dynamics}",
    eprint = "1907.04684",
    archivePrefix = "arXiv",
    primaryClass = "hep-ph",
    doi = "10.1103/PhysRevD.101.091502",
    journal = "Phys. Rev. D",
    volume = "101",
    number = "9",
    pages = "091502",
    year = "2020"
}

@article{Lin:2019qiv,
    author = "Lin, Yong-Hui and Zou, Bing-Song",
    title = "{Strong decays of the latest LHCb pentaquark candidates in hadronic molecule pictures}",
    eprint = "1908.05309",
    archivePrefix = "arXiv",
    primaryClass = "hep-ph",
    doi = "10.1103/PhysRevD.100.056005",
    journal = "Phys. Rev. D",
    volume = "100",
    number = "5",
    pages = "056005",
    year = "2019"
}

@article{Gutsche:2019mkg,
    author = "Gutsche, Thomas and Lyubovitskij, Valery E.",
    title = "{Structure and decays of hidden heavy pentaquarks}",
    eprint = "1910.03984",
    archivePrefix = "arXiv",
    primaryClass = "hep-ph",
    doi = "10.1103/PhysRevD.100.094031",
    journal = "Phys. Rev. D",
    volume = "100",
    number = "9",
    pages = "094031",
    year = "2019"
}

@article{Burns:2019iih,
    author = "Burns, T. J. and Swanson, E. S.",
    title = "{Molecular interpretation of the $P_c$(4440) and $P_c$(4457) states}",
    eprint = "1908.03528",
    archivePrefix = "arXiv",
    primaryClass = "hep-ph",
    doi = "10.1103/PhysRevD.100.114033",
    journal = "Phys. Rev. D",
    volume = "100",
    number = "11",
    pages = "114033",
    year = "2019"
}

@article{Zhu:2019iwm,
    author = "Zhu, Ruilin and Liu, Xuejie and Huang, Hongxia and Qiao, Cong-Feng",
    title = "{Analyzing doubly heavy tetra- and penta-quark states by variational method}",
    eprint = "1904.10285",
    archivePrefix = "arXiv",
    primaryClass = "hep-ph",
    doi = "10.1016/j.physletb.2019.134869",
    journal = "Phys. Lett. B",
    volume = "797",
    pages = "134869",
    year = "2019"
}

@article{Shen:2024nck,
    author = {Shen, Chao-Wei and R\"onchen, Deborah and Mei\ss{}ner, Ulf-G. and Zou, Bing-Song and Wang, Yu-Fei},
    title = "{Exploration of the LHCb $P_c$ states and possible resonances in a unitary coupled-channel model}",
    eprint = "2405.02626",
    archivePrefix = "arXiv",
    primaryClass = "hep-ph",
    doi = "10.1140/epjc/s10052-024-13139-0",
    journal = "Eur. Phys. J. C",
    volume = "84",
    number = "7",
    pages = "764",
    year = "2024"
}

@article{Du:2021fmf,
    author = "Du, Meng-Lin and Baru, Vadim and Guo, Feng-Kun and Hanhart, Christoph and Mei\ss{}ner, Ulf-G. and Oller, Jos\'e A. and Wang, Qian",
    title = "{Revisiting the nature of the P$_{c}$ pentaquarks}",
    eprint = "2102.07159",
    archivePrefix = "arXiv",
    primaryClass = "hep-ph",
    doi = "10.1007/JHEP08(2021)157",
    journal = "JHEP",
    volume = "08",
    pages = "157",
    year = "2021"
}

@article{Wang:2019spc,
    author = "Wang, Guang-Juan and Xiao, Li-Ye and Chen, Rui and Liu, Xiao-Hai and Liu, Xiang and Zhu, Shi-Lin",
    title = "{Probing hidden-charm decay properties of $P_c$ states in a molecular scenario}",
    eprint = "1911.09613",
    archivePrefix = "arXiv",
    primaryClass = "hep-ph",
    doi = "10.1103/PhysRevD.102.036012",
    journal = "Phys. Rev. D",
    volume = "102",
    number = "3",
    pages = "036012",
    year = "2020"
}

@article{Balian:1969sd,
    author = "Balian, R. and Brezin, E.",
    title = "{Angular-momentum reduction of the faddeev equations}",
    doi = "10.1007/BF02710946",
    journal = "Nuovo Cim. B",
    volume = "61",
    pages = "403--410",
    year = "1969"
}

@book{BlattWeisskopf1952,
  author    = {Blatt, John M. and Weisskopf, Victor F.},
  title     = {Theoretical Nuclear Physics},
  publisher = {John Wiley \& Sons},
  address   = {New York},
  year      = {1952},
  isbn      = {9780471080190}
}

@book{Edmonds1957,
  author    = {Edmonds, A. R.},
  title     = {Angular Momentum in Quantum Mechanics},
  publisher = {Princeton University Press},
  address   = {Princeton, New Jersey},
  year      = {1957}
}

@article{Stadler:1991zz,
    author = "Stadler, A. and Glockle, W. and Sauer, P. U.",
    title = "{Faddeev equations with three-nucleon force in momentum space}",
    doi = "10.1103/PhysRevC.44.2319",
    journal = "Phys. Rev. C",
    volume = "44",
    pages = "2319--2327",
    year = "1991"
}

@article{Gloeckle:1982ze,
    author = "Gloeckle, Walter",
    title = "{EFFECTS OF THE TWO PION EXCHANGE THREE NUCLEON FORCE IN THE TRITON AND HE-3}",
    doi = "10.1016/0375-9474(82)90364-5",
    journal = "Nucl. Phys. A",
    volume = "381",
    pages = "343--364",
    year = "1982"
}

@article{Chen:1986pi,
    author = "Chen, C. R. and Payne, G. L. and Friar, James Lewis and Gibson, B. F.",
    title = "{Faddeev Calculations of the (2 $\pi$ - 3n) Force Contribution to the $^{3}$H Binding Energy}",
    doi = "10.1103/PhysRevC.33.1740",
    journal = "Phys. Rev. C",
    volume = "33",
    pages = "1740--1752",
    year = "1986"
}

@article{Chen:1991zza,
    author = "Chen, C. R. and Payne, G. L. and Friar, James Lewis and Gibson, Benjamin F.",
    title = "{Nd zero-energy scattering}",
    doi = "10.1103/PhysRevC.44.50",
    journal = "Phys. Rev. C",
    volume = "44",
    pages = "50--59",
    year = "1991"
}

@article{Lazauskas:2026nlg,
    author = "Lazauskas, R. and Kezerashvili, R. Ya. and Filikhin, I.",
    title = "{Faddeev-Yakubovsky calculations for {\ensuremath{\phi}}-mesic nuclei with the HAL QCD potential}",
    doi = "10.1103/77cb-pbln",
    journal = "Phys. Rev. D",
    volume = "113",
    number = "7",
    pages = "074024",
    year = "2026"
}

@article{Machleidt:2000ge,
    author = "Machleidt, R.",
    title = "{The High precision, charge dependent Bonn nucleon-nucleon potential (CD-Bonn)}",
    eprint = "nucl-th/0006014",
    archivePrefix = "arXiv",
    doi = "10.1103/PhysRevC.63.024001",
    journal = "Phys. Rev. C",
    volume = "63",
    pages = "024001",
    year = "2001"
}

@article{Stoks:1994wp,
    author = "Stoks, V. G. J. and Klomp, R. A. M. and Terheggen, C. P. F. and de Swart, J. J.",
    title = "{Construction of high quality N N potential models}",
    eprint = "nucl-th/9406039",
    archivePrefix = "arXiv",
    reportNumber = "THEF-NYM-93-05",
    doi = "10.1103/PhysRevC.49.2950",
    journal = "Phys. Rev. C",
    volume = "49",
    pages = "2950--2962",
    year = "1994"
}

@article{Wiringa:1994wb,
    author = "Wiringa, Robert B. and Stoks, V. G. J. and Schiavilla, R.",
    title = "{An Accurate nucleon-nucleon potential with charge independence breaking}",
    eprint = "nucl-th/9408016",
    archivePrefix = "arXiv",
    reportNumber = "PHY-7742-TH-94, CEBAF-TH-94-19",
    doi = "10.1103/PhysRevC.51.38",
    journal = "Phys. Rev. C",
    volume = "51",
    pages = "38--51",
    year = "1995"
}

@article{Epelbaum:2014efa,
    author = "Epelbaum, E. and Krebs, H. and Mei{\ss}ner, U. G.",
    title = "{Improved chiral nucleon-nucleon potential up to next-to-next-to-next-to-leading order}",
    eprint = "1412.0142",
    archivePrefix = "arXiv",
    primaryClass = "nucl-th",
    doi = "10.1140/epja/i2015-15053-8",
    journal = "Eur. Phys. J. A",
    volume = "51",
    number = "5",
    pages = "53",
    year = "2015"
}

@article{Malfliet:1968tj,
    author = "Malfliet, R. A. and Tjon, J. A.",
    title = "{Solution of the Faddeev equations for the triton problem using local two particle interactions}",
    doi = "10.1016/0375-9474(69)90775-1",
    journal = "Nucl. Phys. A",
    volume = "127",
    pages = "161--168",
    year = "1969"
}

@article{Friar:1990zza,
    author = "Friar, James Lewis and others",
    title = "{Benchmark solutions for a model three-nucleon scattering problem}",
    doi = "10.1103/PhysRevC.42.1838",
    journal = "Phys. Rev. C",
    volume = "42",
    pages = "1838--1840",
    year = "1990"
}

@article{Leidemann:2012hr,
    author = "Leidemann, Winfried and Orlandini, Giuseppina",
    title = "{Modern Ab Initio Approaches and Applications in Few-Nucleon Physics with A {\ensuremath{>}}= 4}",
    eprint = "1204.4617",
    archivePrefix = "arXiv",
    primaryClass = "nucl-th",
    doi = "10.1016/j.ppnp.2012.09.001",
    journal = "Prog. Part. Nucl. Phys.",
    volume = "68",
    pages = "158--214",
    year = "2013"
}

@article{Navratil:1999pw,
    author = "Navratil, P. and Kamuntavicius, G. P. and Barrett, B. R.",
    title = "{Few nucleon systems in translationally invariant harmonic oscillator basis}",
    eprint = "nucl-th/9907054",
    archivePrefix = "arXiv",
    doi = "10.1103/PhysRevC.61.044001",
    journal = "Phys. Rev. C",
    volume = "61",
    pages = "044001",
    year = "2000"
}

@article{Kievsky:2008es,
    author = "Kievsky, A. and Rosati, S. and Viviani, M. and Marcucci, L. E. and Girlanda, L.",
    title = "{A High-precision variational approach to three- and four-nucleon bound and zero-energy scattering states}",
    eprint = "0805.4688",
    archivePrefix = "arXiv",
    primaryClass = "nucl-th",
    doi = "10.1088/0954-3899/35/6/063101",
    journal = "J. Phys. G",
    volume = "35",
    pages = "063101",
    year = "2008"
}

@article{Gloeckle:1995jg,
    author = "Gloeckle, Walter and Witala, H. and Huber, D. and Kamada, H. and Golak, J.",
    title = "{The Three nucleon continuum: Achievements, challenges and applications}",
    doi = "10.1016/0370-1573(95)00085-2",
    journal = "Phys. Rept.",
    volume = "274",
    pages = "107--285",
    year = "1996"
}

@Book{abramowitz+stegun,
  author    = "Milton Abramowitz and Irene A. Stegun",
  title     = "Handbook of Mathematical Functions with Formulas, Graphs, and Mathematical Tables",
  publisher = "Dover",
  year      = 1964,
  address   = "New York",
  isbn      = "0-486-61272-4"
}

@article{Lazauskas:2026ysy,
    author = "Lazauskas, Rimantas and Kezerashvili, Roman Ya. and Filikhin, Igor",
    title = "{Possible Existence of $^3_ϕ$H, $^4_ϕ$H, $^4_ϕ$He, and $^5_ϕ$He Nuclei}",
    eprint = "2601.14572",
    archivePrefix = "arXiv",
    primaryClass = "nucl-th",
    doi = "10.1016/j.physletb.2026.140496",
    journal = "Phys. Lett. B",
    volume = "877",
    pages = "140496",
    year = "2026"
}

@article{Yokota:2013sfa,
    author = "Yokota, Akira and Hiyama, Emiko and Oka, Makoto",
    title = "{Possible existence of charmonium{\textendash}nucleus bound states}",
    eprint = "1308.6102",
    archivePrefix = "arXiv",
    primaryClass = "nucl-th",
    doi = "10.1093/ptep/ptt087",
    journal = "PTEP",
    volume = "2013",
    number = "11",
    pages = "113D01",
    year = "2013"
}

@book{BurdenFaires2016,
    author    = {Burden, Richard L. and Faires, J. Douglas and Burden, Annette M.},
    title     = {Numerical Analysis},
    edition   = {10th},
    publisher = {Cengage Learning},
    year      = {2016},
    isbn      = {9781305253667}
}

@book{PressTeukolsky2007,
    author    = {Press, William H. and Teukolsky, Saul A. and Vetterling, William T. and Flannery, Brian P.},
    title     = {Numerical Recipes: The Art of Scientific Computing},
    edition   = {3rd},
    publisher = {Cambridge University Press},
    year      = {2007},
    isbn      = {9780521880688}
}

@article{Strakovsky:2026qkv,
    author = "Strakovsky, Igor I. and Briscoe, William J. and Ahn, Jung Keun and Ryskin, Misha G. and Schmidt, Axel",
    title = "{JLab and J-PARC measurements for J/{\ensuremath{\psi}} production at threshold}",
    eprint = "2603.09622",
    archivePrefix = "arXiv",
    primaryClass = "hep-ph",
    doi = "10.1103/3g18-qv5t",
    journal = "Phys. Rev. D",
    volume = "113",
    number = "11",
    pages = "114023",
    year = "2026"
}

@article{Wang:2015jsa,
    author = "Wang, Qian and Liu, Xiao-Hai and Zhao, Qiang",
    title = "{Photoproduction of hidden charm pentaquark states $P_c^+(4380)$ and $P_c^+(4450)$}",
    eprint = "1508.00339",
    archivePrefix = "arXiv",
    primaryClass = "hep-ph",
    doi = "10.1103/PhysRevD.92.034022",
    journal = "Phys. Rev. D",
    volume = "92",
    pages = "034022",
    year = "2015"
}

@article{Kubarovsky:2015aaa,
    author = "Kubarovsky, V. and Voloshin, M. B.",
    title = "{Formation of hidden-charm pentaquarks in photon-nucleon collisions}",
    eprint = "1508.00888",
    archivePrefix = "arXiv",
    primaryClass = "hep-ph",
    reportNumber = "FTPI-MINN-15-38, UMN-TH-3447-15, JLAB-PHY-15-2131",
    doi = "10.1103/PhysRevD.92.031502",
    journal = "Phys. Rev. D",
    volume = "92",
    number = "3",
    pages = "031502",
    year = "2015"
}

@article{Karliner:2015voa,
    author = "Karliner, Marek and Rosner, Jonathan L.",
    title = "{Photoproduction of Exotic Baryon Resonances}",
    eprint = "1508.01496",
    archivePrefix = "arXiv",
    primaryClass = "hep-ph",
    reportNumber = "EFI-15-25, TAUP-2998-15",
    doi = "10.1016/j.physletb.2015.11.068",
    journal = "Phys. Lett. B",
    volume = "752",
    pages = "329--332",
    year = "2016"
}

@article{HillerBlin:2016odx,
    author = "Hiller Blin, A. N. and Fern\'andez-Ram\'\i{}rez, C. and Jackura, A. and Mathieu, V. and Mokeev, V. I. and Pilloni, A. and Szczepaniak, A. P.",
    title = "{Studying the P$_c$(4450) resonance in J/$\psi$ photoproduction off protons}",
    eprint = "1606.08912",
    archivePrefix = "arXiv",
    primaryClass = "hep-ph",
    reportNumber = "JLAB-THY-16-2277",
    doi = "10.1103/PhysRevD.94.034002",
    journal = "Phys. Rev. D",
    volume = "94",
    number = "3",
    pages = "034002",
    year = "2016"
}

@article{Winney:2019edt,
    author = "Winney, Daniel and Fanelli, Cristiano and Pilloni, Alessandro and Hiller Blin, Astrid N. and Fern\'andez-Ram\'\i{}rez, Cesar and Albaladejo, Miguel and Mathieu, Vincent and Mokeev, Victor I. and Szczepaniak, Adam P.",
    collaboration = "JPAC",
    title = "{Double polarization observables in pentaquark photoproduction}",
    eprint = "1907.09393",
    archivePrefix = "arXiv",
    primaryClass = "hep-ph",
    reportNumber = "JLAB-THY-19-3004",
    doi = "10.1103/PhysRevD.100.034019",
    journal = "Phys. Rev. D",
    volume = "100",
    number = "3",
    pages = "034019",
    year = "2019"
}

@article{Wang:2019krd,
    author = "Wang, Xiao-Yun and Chen, Xu-Rong and He, Jun",
    title = "{Possibility to study pentaquark states $P_{c}(4312), P_{c}(4440)$, and $P_{c}(4457)$ in $\gamma p\rightarrow J/\psi p$ reaction}",
    eprint = "1904.11706",
    archivePrefix = "arXiv",
    primaryClass = "hep-ph",
    doi = "10.1103/PhysRevD.99.114007",
    journal = "Phys. Rev. D",
    volume = "99",
    number = "11",
    pages = "114007",
    year = "2019"
}

@article{Wu:2019adv,
    author = "Wu, Jia-Jun and Lee, T. -S. H. and Zou, Bing-Song",
    title = "{Nucleon resonances with hidden charm in \ensuremath{\gamma}p reactions}",
    eprint = "1906.05375",
    archivePrefix = "arXiv",
    primaryClass = "nucl-th",
    doi = "10.1103/PhysRevC.100.035206",
    journal = "Phys. Rev. C",
    volume = "100",
    number = "3",
    pages = "035206",
    year = "2019"
}

@article{Paryev:2022wov,
    author = "Paryev, E. Ya.",
    title = "{Towards clarifying the possibility of observation of the LHCb hidden-charm pentaquarks Pc+(4312), Pc+(4337), Pc+(4440) and Pc+(4457) in near-threshold charmonium photoproduction off protons and nuclei}",
    eprint = "2211.16037",
    archivePrefix = "arXiv",
    primaryClass = "hep-ph",
    doi = "10.1016/j.nuclphysa.2022.122562",
    journal = "Nucl. Phys. A",
    volume = "1029",
    pages = "122562",
    year = "2023"
}

@article{Cao:2019kst,
    author = "Cao, Xu and Dai, Jian-ping",
    title = "{Confronting pentaquark photoproduction with new LHCb observations}",
    eprint = "1904.06015",
    archivePrefix = "arXiv",
    primaryClass = "hep-ph",
    doi = "10.1103/PhysRevD.100.054033",
    journal = "Phys. Rev. D",
    volume = "100",
    number = "5",
    pages = "054033",
    year = "2019"
}

@article{Strakovsky:2023kqu,
    author = "Strakovsky, Igor and Briscoe, William J. and Chudakov, Eugene and Larin, Ilya and Pentchev, Lubomir and Schmidt, Axel and Workman, Ronald L.",
    title = "{Plausibility of the LHCb Pc(4312)+ in the GlueX \ensuremath{\gamma}p\textrightarrow{}J/\ensuremath{\psi}p total cross sections}",
    eprint = "2304.04924",
    archivePrefix = "arXiv",
    primaryClass = "hep-ph",
    reportNumber = "JLAB-PHY-23-3888",
    doi = "10.1103/PhysRevC.108.015202",
    journal = "Phys. Rev. C",
    volume = "108",
    number = "1",
    pages = "015202",
    year = "2023"
}

@article{Clymton:2024fbf,
    author = "Clymton, Samson and Kim, Hyun-Chul and Mart, Terry",
    title = "{Production mechanism of the hidden charm pentaquark states $P_{c\bar{c}}$}",
    eprint = "2408.04166",
    archivePrefix = "arXiv",
    primaryClass = "hep-ph",
    reportNumber = "INHA-NTG-05/2024",
    month = "8",
    year = "2024"
}

@article{Duan:2024hby,
    author = "Duan, Ming-Xiao and Gong, Chang and Qiu, Lin and Zhao, Qiang",
    title = "{$J/\psi$ photoproduction near threshold and signals for the hidden charm pentaquarks}",
    eprint = "2409.10364",
    archivePrefix = "arXiv",
    primaryClass = "hep-ph",
    month = "9",
    year = "2024"
}

@article{Clymton:2026kka,
    author = "Clymton, Samson and Kim, Sang-Ho and Kim, Hyun-Chul",
    title = "{Nonexistence of hidden-charm pentaquarks in $J/ψ$ photoproduction}",
    eprint = "2606.11808",
    archivePrefix = "arXiv",
    primaryClass = "hep-ph",
    reportNumber = "INHA-NTG-05/2026",
    month = "6",
    year = "2026"
}

@article{Huang:2005gw,
    author = "Huang, F. and Zhang, Z. Y. and Yu, Y. W.",
    title = "{N phi state in chiral quark model}",
    eprint = "nucl-th/0512079",
    archivePrefix = "arXiv",
    doi = "10.1103/PhysRevC.73.025207",
    journal = "Phys. Rev. C",
    volume = "73",
    pages = "025207",
    year = "2006"
}

@article{He:2018plt,
    author = "He, Jun and Huang, Hongxia and Chen, Dian-Yong and Zhu, Xinmei",
    title = "{Hidden-strange molecular states and the N{\ensuremath{\phi}} bound states via a QCD van der Waals force}",
    eprint = "1804.09383",
    archivePrefix = "arXiv",
    primaryClass = "hep-ph",
    doi = "10.1103/PhysRevD.98.094019",
    journal = "Phys. Rev. D",
    volume = "98",
    number = "9",
    pages = "094019",
    year = "2018"
}

@article{Gao:2017hya,
    author = "Gao, Haiyan and Huang, Hongxia and Liu, Tianbo and Ping, Jialun and Wang, Fan and Zhao, Zhiwen",
    title = "{Search for a hidden strange baryon-meson bound state from {\ensuremath{\phi}} production in a nuclear medium}",
    eprint = "1701.03210",
    archivePrefix = "arXiv",
    primaryClass = "hep-ph",
    doi = "10.1103/PhysRevC.95.055202",
    journal = "Phys. Rev. C",
    volume = "95",
    number = "5",
    pages = "055202",
    year = "2017"
}

@article{Ishikawa:2004id,
    author = "Ishikawa, T. and others",
    title = "{phi photo-production from Li, C, Al, and Cu nuclei at E(gamma) = 1.5-GeV to 2.4-GeV}",
    eprint = "nucl-ex/0411016",
    archivePrefix = "arXiv",
    doi = "10.1016/j.physletb.2005.01.023",
    journal = "Phys. Lett. B",
    volume = "608",
    pages = "215--222",
    year = "2005"
}

@article{CLAS:2010pxs,
    author = "Wood, M. H. and others",
    collaboration = "CLAS",
    title = "{Absorption of the $\omega$ and $\phi$ Mesons in Nuclei}",
    eprint = "1006.3361",
    archivePrefix = "arXiv",
    primaryClass = "nucl-ex",
    reportNumber = "JLAB-PHY-10-1156",
    doi = "10.1103/PhysRevLett.105.112301",
    journal = "Phys. Rev. Lett.",
    volume = "105",
    pages = "112301",
    year = "2010"
}

@article{Polyanskiy:2010tj,
    author = "Polyanskiy, A. and others",
    title = "{Measurement of the in-medium phi-meson width in proton-nucleus collisions}",
    eprint = "1008.0232",
    archivePrefix = "arXiv",
    primaryClass = "nucl-ex",
    doi = "10.1016/j.physletb.2010.10.050",
    journal = "Phys. Lett. B",
    volume = "695",
    pages = "74--77",
    year = "2011"
}

@article{KEK-PS-E325:2005wbm,
    author = "Muto, R. and others",
    collaboration = "KEK-PS-E325",
    title = "{Evidence for in-medium modification of the phi meson at normal nuclear density}",
    eprint = "nucl-ex/0511019",
    archivePrefix = "arXiv",
    doi = "10.1103/PhysRevLett.98.042501",
    journal = "Phys. Rev. Lett.",
    volume = "98",
    pages = "042501",
    year = "2007"
}

@article{Dey:2014tfa,
    author = "Dey, B. and Meyer, C. A. and Bellis, M. and Williams, M",
    collaboration = "CLAS",
    title = "{Data analysis techniques, differential cross sections, and spin density matrix elements for the reaction $\gamma p \rightarrow \phi p$}",
    eprint = "1403.2110",
    archivePrefix = "arXiv",
    primaryClass = "nucl-ex",
    doi = "10.1103/PhysRevC.89.055208",
    journal = "Phys. Rev. C",
    volume = "89",
    number = "5",
    pages = "055208",
    year = "2014",
    note = "[Addendum: Phys.Rev.C 90, 019901 (2014)]"
}

@article{Strakovsky:2020uqs,
    author = "Strakovsky, Igor I. and Pentchev, Lubomir and Titov, Alexander",
    title = "{Comparative analysis of $\omega p$, $\phi p$, and $J/\psi p$ scattering lengths from A2, CLAS, and GlueX threshold measurements}",
    eprint = "2001.08851",
    archivePrefix = "arXiv",
    primaryClass = "hep-ph",
    doi = "10.1103/PhysRevC.101.045201",
    journal = "Phys. Rev. C",
    volume = "101",
    number = "4",
    pages = "045201",
    year = "2020"
}

@article{Laverne:1973kgl,
    author = "Laverne, A. and Gignoux, C.",
    title = "{A detailed analysis of 3 H from Faddeev equations in configuration space}",
    doi = "10.1016/0375-9474(73)90366-7",
    journal = "Nucl. Phys. A",
    volume = "203",
    pages = "597--608",
    year = "1973"
}

@article{Payne:1980ap,
    author = "Payne, G. L. and Friar, James Lewis and Gibson, B. F. and Afnan, I. R.",
    title = "{CONFIGURATION SPACE FADDEEV CALCULATIONS. 1. TRITON GROUND STATE PROPERTIES}",
    reportNumber = "LA-UR-80-67",
    doi = "10.1103/PhysRevC.22.823",
    journal = "Phys. Rev. C",
    volume = "22",
    pages = "823",
    year = "1980"
}

@article{Payne:1980aq,
    author = "Payne, G. and Gibson, B. F. and Friar, James Lewis",
    title = "{CONFIGURATION SPACE FADDEEV CALCULATIONS. 2. TRINUCLEON COULOMB ENERGY}",
    reportNumber = "LA-UR-80-52",
    doi = "10.1103/PhysRevC.22.832",
    journal = "Phys. Rev. C",
    volume = "22",
    pages = "832",
    year = "1980"
}

@article{Haftel:1970zz,
    author = "Haftel, Michael I. and Tabakin, Frank",
    title = "{NUCLEAR SATURATION AND THE SMOOTHNESS OF NUCLEON-NUCLEON POTENTIALS}",
    reportNumber = "PRINT-70-1884",
    doi = "10.1016/0375-9474(70)90047-3",
    journal = "Nucl. Phys. A",
    volume = "158",
    pages = "1--42",
    year = "1970"
}
\end{document}